\documentclass[a4paper]{article}

\newcommand{\softwareName}{\textbf{InferenceMAP}~}

\newcommand{\macFile}{InferenceMAP for Mac.dmg}
\newcommand{\macExecutable}{InferenceMAP}
\newcommand*{\TitleParbox}[1]{\parbox[c]{2cm}{\raggedright #1}}
\newcommand*{\TitleParboxA}[1]{\parbox[c]{4.5cm}{\raggedright #1}}

\usepackage{fullpage}
\usepackage{geometry}
\usepackage[T1]{fontenc}
\usepackage[dvips]{color}
\usepackage{soul}
\usepackage{multicol}
\usepackage{eurosym}
\usepackage{transparent}
\usepackage{hyperref}
\hypersetup{
	pdftitle = \softwareName User Manual,
	pdfauthor = Mohamed El Beheiry,
	pdfborder = {0 0 0},
}
\usepackage{fancyhdr}
\usepackage{lastpage}
\usepackage{graphicx}
\usepackage{array}
\usepackage{latexsym}
\usepackage{siunitx}
\usepackage{amsmath} 
\usepackage{esint}
\usepackage{caption}
\usepackage{longtable}
\usepackage{subfig}
\usepackage{pdfpages}

\definecolor{gray}{RGB}{190,190,190}

\title{\textbf{\softwareName User Manual}}
\author{Mohamed El Beheiry}
\date{2014}

\begin{document}



# InferenceMAP: Mapping of Single-Molecule Dynamics with Bayesian Inference

Mohamed El Beheiry[1,2], Maxime Dahan[1] and Jean-Baptiste Masson[2,3]

[1] Physico-Chimie Curie, Institut Curie, CNRS UMR 168, Université Pierre et Marie Curie-Paris 6, 26 rue d'Ulm 75248 Paris, France
[2] Institut Pasteur, Physics of Biological Systems, CNRS UMR 3525, 25/28 rue du Docteur Roux, 75015 Paris, France
[3] Janelia Research Campus, Howard Hughes Medical Institute, Ashburn, VA 20147, USA

Single-particle tracking (SPT) grants unprecedented insight into cellular function at the molecular scale [1]. Throughout the cell, the movement of single-molecules is generally heterogeneous and complex. Hence, there is an imperative to understand the multi-scale nature of single-molecule dynamics in biological systems. We have previously shown that with high-density SPT, spatial maps of the parameters that dictate molecule motion can be generated to intricately describe cellular environments [2,3,4]. To date, however, there exist no publically available tools that reconcile trajectory data to generate the aforementioned maps. We address this void in the SPT community with InferenceMAP: an interactive software package that uses a powerful Bayesian method to map the dynamic cellular space experienced by individual biomolecules (**Supplementary Software**).

High-density SPT methods, such as sptPALM [5] and uPAINT [6], capture thousands of molecule trajectories in a few minutes of acquisition at high spatiotemporal resolution. By using one of the many available particle-tracking algorithms [7], trajectories are reconstituted and accepted as input to InferenceMAP. This massive trajectory data is thereafter treated with a Bayesian inference algorithm that, notably, imposes no constraints on trajectory lengths [2,3,4]. Our algorithm is compatible with different models of single-molecule motion, including hopping diffusion (**Fig. 1a**), active processes, confinement, and interaction energy-driven systems (**Fig. 1b**). Model-specific physical processes are distinguished and mapped, revealing rich landscapes of molecule dynamics.

Generating dynamical maps from single-particle trajectories is critically dependent on the meshing utilized. As local diffusivities may vary by orders of magnitude in a few hundred nanometers, meshes should locally adapt to match the characteristic size of molecule displacements. To this end, InferenceMAP offers adaptive meshing techniques that users may tune to fit the spatial organization of their single-molecule trajectories (**Fig. 1a,c**). In each zone of a mesh, dynamic parameters are inferred to give rise to a parameter landscape (**Supplementary Information**).

Calculations can be performed in an automated fashion irrespective of the biological system, or parameters may be carefully adjusted to conform to desired mapping resolution and optimization constraints. Additionally, beforehand knowledge of the biological system can be incorporated in calculations via user-defined prior probabilities (**Supplementary Information**). Furthermore, a randomized optimization algorithm is available for exceptionally large problems, permitting mapping of entire cells such as in **Fig. 1b**. Highly-localized analysis in sub-regions is easily performed through a custom-selection macro. Resulting inferred data is exportable in image and ASCII formats.

InferenceMAP offers a host of features to address a major concern of the single-particle tracking community: it reveals the parameters that dictate the motion of molecules. Moreover, in the burgeoning age of big data experimental biophysics, it is an important contribution that extracts sensible results from otherwise dense and complicated observations using a robust Bayesian method. InferenceMAP is controlled with a user-friendly interface and is compatible with Mac OS X and Windows. The software is freely available for academic use (source code is available upon signature of a Material Transfer Agreement), and latest versions may be downloaded from http://umr168.curie.fr/en/research-groups/locco/software and from http://www.pasteur.fr/en/research/genomes-genetics/units-groups/jean-baptiste-masson. The authors request acknowledgment of the use of InferenceMAP in published works. Prior to publication, the software will be accessible through an email request to Mohamed El Beheiry (mohamed.el-beheiry@curie.fr) or to Jean-Baptiste Masson (jbmasson@pasteur.fr).

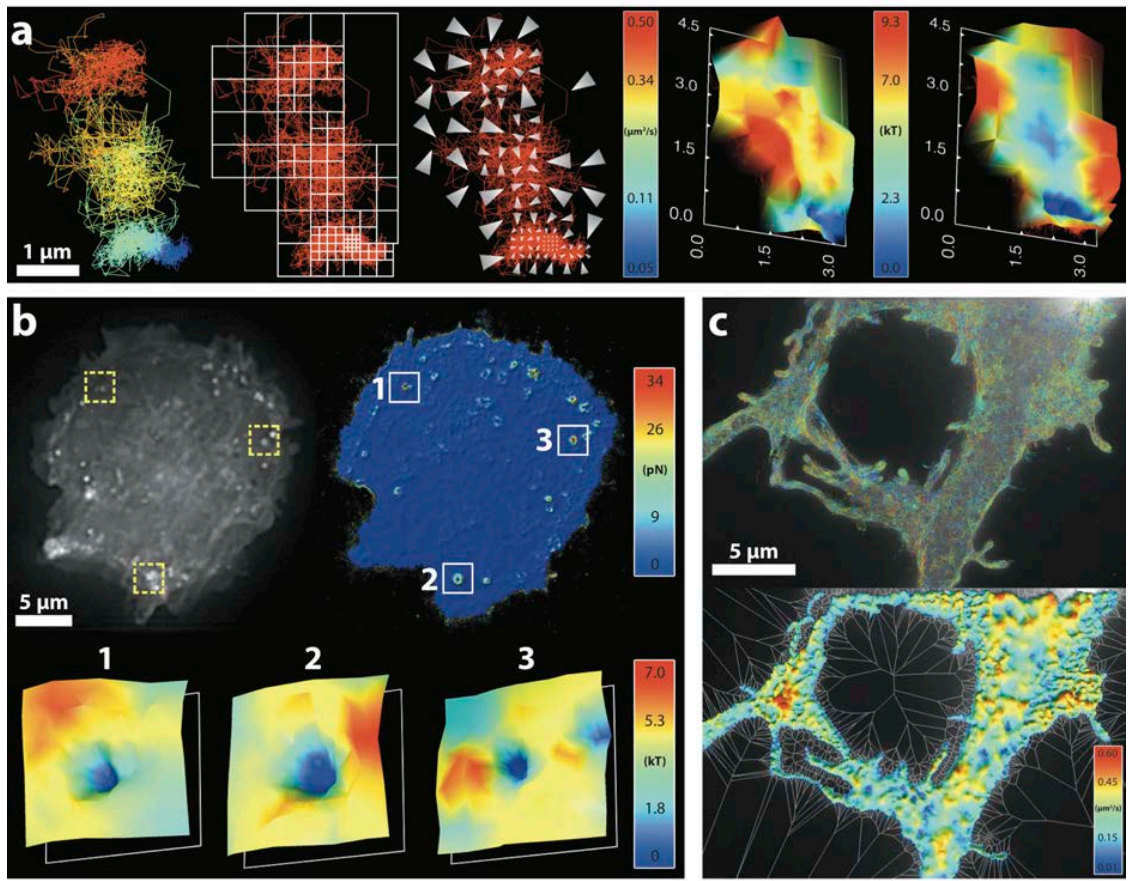

**Figure 1 Caption**

Fig 1. Various uses of InferenceMAP on experimental single-particle trajectories. Shown are various ways to meshing the area explored by trajectories and to represent inferred dynamical parameters (diffusivity, force components, and interaction potentials).

a. Typical analysis workflow in InferenceMAP, from left to right. First panel: trajectory of an $\varepsilon$-toxin receptor tagged to an amine-coated lanthanide oxide nanoparticle on an MDCK cell hopping between different lipid rafts [3]. The color code is associated to time, from blue (beginning) to red (ending) for a duration of 290 s. Second panel: same trajectory in red, overlaid with a quad-tree mesh. Note the multi-scale nature of the mesh, which depends on both the number of points and the local diffusivity. Third panel: force map (directional bias) acting on the receptor in its various confinement areas, indicated by white arrows. Fourth panel: Three-dimensional landscape view of diffusivity. Fifth panel: Landscape view of interaction energy map, revealing three membrane confinement domains.

b. Diffusivity map of glycine receptor construct (pHluorin-TMD-$\beta$-Loop 400AA [4]) in a mouse hippocampal neuron from a uPAINT measurement with an anti-GFP antibody coupled to atto647N. Top panel: 9,453 trajectories of the GlyR construct overlaid to an ensemble (GFP) fluorescence image of receptors. Color code distinguishes different trajectories. Bottom panel: three-dimensional diffusivity map plotted overtop the employed Voronoi tessellated mesh. The Voronoi mesh allows for efficient mapping of cells with complex geometries, and clearly shows local variability of the diffusivity.

c. Whole-cell interaction energy map of glycine receptor construct (Dendra2-TMD-ß-Loop WT) with gephyrin clusters using TIRF microscopy. Top-left panel: ensemble fluorescence image of Cerulean-Gephyrin clusters (yellow boxes) in a COS-7 cell. Top-right panel: Three-dimensional map of force amplitudes at the basal membrane of the cell. Note that the map is mostly flat, except in highly-localized regions colocalized with gephyrin clusters (white boxes). Interaction energy wells corresponding to the boxed regions are indicated in the bottom three panels.



\begin{titlepage}

\begin{center}

\vspace*{5 cm}

\includegraphics[width=200pt]{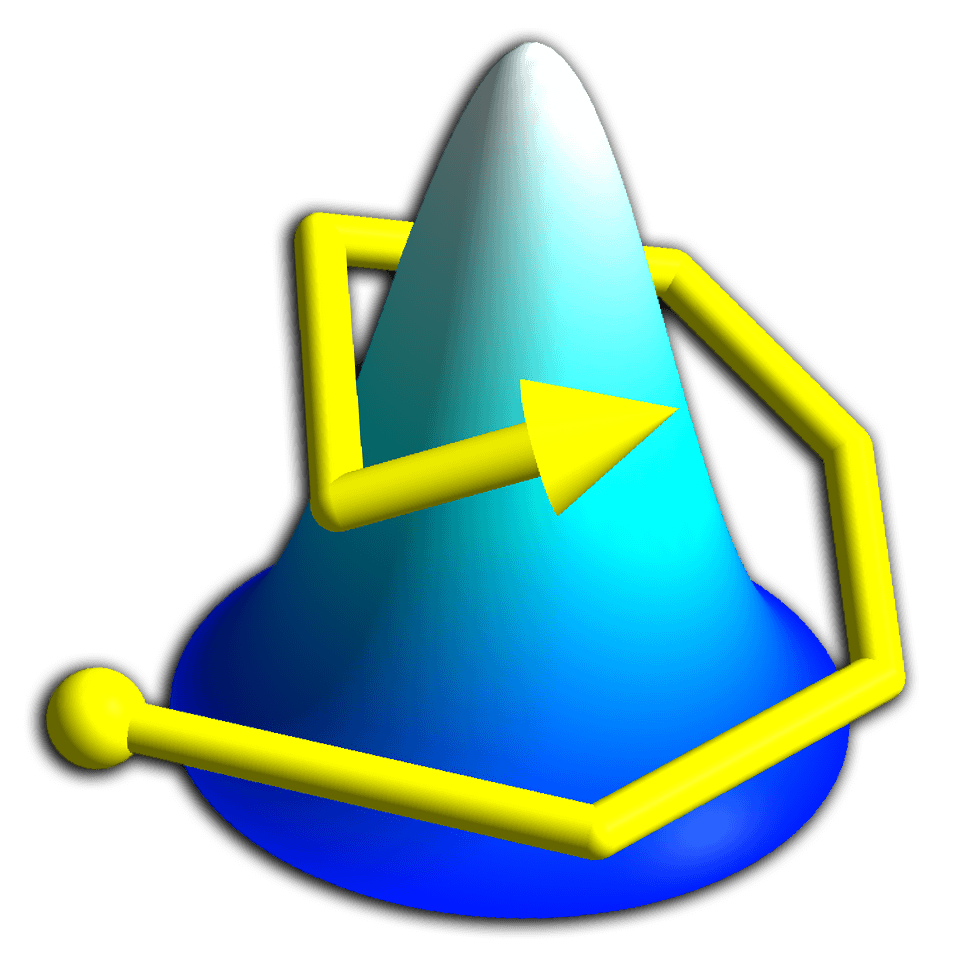}~\\[1cm]

\textbf{\LARGE \softwareName User Manual}\\
\vspace*{1 cm}
\textbf{\Large Version 1.0}

\vspace*{2 cm}

\textbf{Mohamed El Beheiry}\\
Institut Curie - Centre de Recherche\\
Laboratoire Physico-Chimie Curie (UMR168)\\
26, rue d'Ulm \\
75248 Paris, France\\
\href{mailto:mohamed.el-beheiry@curie.fr}{mohamed.elbeheiry@gmail.com} \\
\vspace*{0.2 cm}
\textbf{Jean-Baptiste Masson}\\
Institut Pasteur\\
Physics of Biological Systems\\
25/28 rue du Docteur Roux\\
75015 Paris , France\\
\href{mailto:jbmasson@pasteur.fr}{jbmasson@pasteur.fr} \\

\vfill

\large{\softwareName is registered with the \emph{Agency for the Protection of Programs} (APP) under reference \textbf{FR.001.350042.000.S.P.2014.000.20700}}

\end{center}
\end{titlepage}

\newpage
\tableofcontents

\newpage

\section{Preliminary Remarks}
\label{sec:preliminary_remarks}

The trajectory inference technique utilized by \softwareName (first described in \cite{masson2009}) is used to study the dynamics of single molecules. Specifically, it generates maps of the dynamic parameters that dictate the motion of single molecules. These parameters may include the diffusion, forces (directional biases), interaction (potential) energies, and drift. Mapping these spatially-dependent parameters is of great interest for a variety of studies. First, dynamical maps can be used to decipher physical interactions a molecule may have with its surroundings. Secondly, generated maps reliably distinguish physical parameters from one another. Finally, it is a means of probing the strength of single molecule interactions. \\

To date, this technique has been used to study the dynamics of single molecules in many biological contexts, these include:

\begin{itemize}
	\item Neurotransmitter receptors \cite{masson2014}
	\item Membrane microdomains \cite{turkcan2012_2}
	\item Transcription factors
	\item Viral capsid fusing proteins
	\item Proteins on unilamellar vesicles
\end{itemize} 

\subsection{Models of Motion}

The widespread applicability of \softwareName is largely due to the different models of single molecule motion it supports. Below the models are briefly described (a full description is available in Section \ref{sec:inference}).

\begin{itemize}
	\item \textbf{Diffusion Only.} Solely the diffusivity is estimated from the trajectories.
	\item \textbf{Diffusion and Force.} Local force components in addition to the diffusion are inferred, with an option to estimate interaction energies.
	\item \textbf{Diffusion and Drift.} Local drift (speed) in addition to the diffusion are inferred. This model is applicable to systems possessing active processes, where forces may not be conservative.
	\item \textbf{Diffusion and Potential.} Potential (interaction) energy, force, and diffusion components are estimated from trajectories.
	\item \textbf{Polynomial Potential.} Potential (interaction) energy, force, and diffusion components are estimated from trajectories. This model is applicable to small regions where there is trajectory confinement.
\end{itemize}

\subsection{Constraints}

Users should be aware of constraints regarding the single molecule trajectory inference technique implemented in \textbf{InferenceMAP}. With its various inference modes and features, \softwareName handles trajectory motion that can be modeled by an \emph{overdamped Langevin} equation. This model is a good approximation to memoryless (Markovian) motion, which single molecules typically exhibit. \\

Below, potential constraints to the use of \softwareName are listed.

\begin{itemize}
	\item \textbf{Limited Localization Density}. If datasets have a low number of trajectory points (i.e. localizations), the precision of inferred parameters may be impeded. A possible way to get around this issue is to reduce the spatial resolution of the mesh in which parameters are inferred. General rules for localization numbers are given in Section \ref{sec:meshing}.
	\item \textbf{Timescale of Dynamics}. Users should be aware of the general timescale of dynamics within their system (e.g. the duration of an interaction or a transport event). \softwareName allows trajectories to be temporally windowed to accommodate different event durations.
	\item \textbf{Viscoelastic Motion}. Viscoelastic motion is not accurately described by the overdamped Langevin equation, in which case \softwareName is not advised for estimating dynamics. Such motion is observed in the motion of large cytosolic vesicles, for example.
	\item \textbf{Immobile Trajectories}. The overdamped Langevin equation does not describe entirely immobilized trajectories. In cases where trajectories have mixed mobile and immobile populations, it is advised to segregate populations prior to analysis with \softwareName.
\end{itemize}

\subsection{Tracking Software}

As \softwareName is used downstream from the single molecule tracking step (it takes trajectories as input data), a tracking software is needed to reconstruct trajectory translocations from acquired microscopy images. A few of the freely available tracking software tools are listed in the references below.

\begin{itemize}
	\item \textbf{u-track} Jaqaman et al., \emph{Nature Methods} 5, pp. 695--702 (2008)
	\item \textbf{MTT} Serg\'{e} et al., \emph{Nature Methods} 5, pp. 687--694 (2008)
	\item \textbf{Various} Chenouard et al., \emph{Nature Methods} 11, pp. 281--289 (2014)
\end{itemize}

\newpage
\section{Installation \& Execution}
\label{sec:installation}
\subsection{Windows (XP \& 7)}

\softwareName has been tested on the \textbf{Windows XP}, \textbf{Windows 7}, and \textbf{Windows 8} operating systems. The installation instructions are as follows:
\begin{enumerate}
	\item Assure that the \textbf{Microsoft Visual C++ 2010 Redistributable Package} is installed. It can be found here: \\ \textbf{\href{http://www.microsoft.com/en-us/download/details.aspx?id=5555}{http://www.microsoft.com/en-us/download/details.aspx?id=5555}}.
	\item Double-click the install package \textbf{InferenceMAP Windows.zip} and follow the on-screen installation instructions.
	\item Double-click the \textbf{InferenceMAP.exe} in the install directory to start the program.
\end{enumerate}
	The execution of \softwareName is coupled with that of a Windows \textbf{Command Prompt}. The purpose of this is to display the progress in potentially lengthy calculations.

\subsubsection{Known Issues}
Reported issues with the \textbf{Windows} version of \softwareName may potentially be circumvented trying the following steps:

\begin{enumerate}
	\item Place the contents of the ViSP installation package directly in your HD path (e.g. place them in a directory such as \textbf{C:/InferenceMAP/}).
	\item Try running ViSP after renaming the .dll files in the installation directory (e.g. rename \textbf{opengl32.dll} to \textbf{opengl32\_temp.dll} and see if it runs).
	\item Assure that you have the latest drivers for your graphics card installed.
\end{enumerate}

\subsection{Mac OS X}
\softwareName has been tested on versions \textbf{10.6.x} (Snow Leopard), \textbf{10.7.x} (Lion), \textbf{10.8.x} (Mountain Lion), and \textbf{10.9.x} (Mavericks) of the \textbf{Mac OS X} operating system. The installation instructions are as follows:

\begin{enumerate}
	\item Mount \textbf{\macFile} by double-clicking it.
	\item Double click the \textbf{InferenceMAP.pkg} file and follow the on-screen installation instructions.
	\item Click the \textbf{\macExecutable} file in the chosen installation directory to start the program.
\end{enumerate}

\subsubsection{Known Issues}
Reported issues with the \textbf{Mac} version of \softwareName may potentially be circumvented trying the following steps:

\begin{enumerate}
	\item If opening \textbf{InferenceMAP.app} presents the message: \\
		\textbf{``InferenceMAP'' can't be opened because it is from an unidentified developer.}\\
	Instead, right-click \textbf{InferenceMAP.app} and select \textbf{Open}. The following the message will appear: \\
		\textbf{``InferenceMAP'' is from an unidentified developer. Are you sure you want to open it?} \\
	Select \textbf{Open} to launch \softwareName.
	\item If opening \textbf{InferenceMAP.app} presents the message: \\
		\textbf{``InferenceMAP.app'' is damaged and can't be opened. You should move it to the Trash.} \\
	Go to \textbf{System Preferences}, select \textbf{Security \& Privacy}. Choose the \textbf{General} tab. Click the lock in the bottom-left corner, after which you will be request to enter your user credentials and password. Under the \textbf{Allow apps downloaded from:}, select \textbf{Anywhere}. Now, when launching \textbf{InferenceMAP.app} a window will appear showing \textbf{``InferenceMAP.app'' is an application downloaded from the Internet. Are you sure you want to open it?}, select \textbf{Open}.
\end{enumerate}


\clearpage

\section{File Formats}
\label{sec:file_formats}

\softwareName supports two ASCII (text) file formats for single-particle trajectory information: 

\begin{enumerate}
	\item Columns are tab-delimited
	\item All geometric units are in micrometers [\SI{}{\micro\meter}]
	\item Temporal information is given in seconds [s]
\end{enumerate}

Files are opened by selecting \textbf{Open} in the \textbf{File} menu. Up to 10 files may be opened in a session, tabs for which will appear in the \textbf{File Tabs} panel of the main interface (see Section \ref{sec:main_interface}).

\subsection{x y t File (.xyt)}
This format is for \emph{single} trajectories. The user has the option of selecting multiple files of this type in the \textbf{File Open} dialog box (each one corresponding to a single trajectory). Columns are defined as follows: \\

\begin{center}
\begin{tabular}{c|c|c}
  \textbf{Column 1} & \textbf{Column 2} & \textbf{Column 3} \\
  \hline
  x-Coordinate [\SI{}{\micro\meter}] & y-Coordinate [\SI{}{\micro\meter}] & Time Stamp [s] \\
\end{tabular}
\end{center}

\subsection{tr x y t File (.trxyt)}
\label{sec:2dlp_file}
This format is for \emph{multiple} trajectories. Columns are defined as follows: \\

\begin{center}
\begin{tabular}{c|c|c|c|c|c}
  \textbf{Column 1} & \textbf{Column 2} & \textbf{Column 3} & \textbf{Column 4} \\
  \hline
  Trajectory Number & x-Coordinate [\SI{}{\micro\meter}] & y-Coordinate [\SI{}{\micro\meter}] & Time Stamp [s] \\
\end{tabular}
\end{center}

\clearpage

\section{Main Interface}
\label{sec:main_interface}

\begin{figure}[h!]
\center{\includegraphics[width=\textwidth]{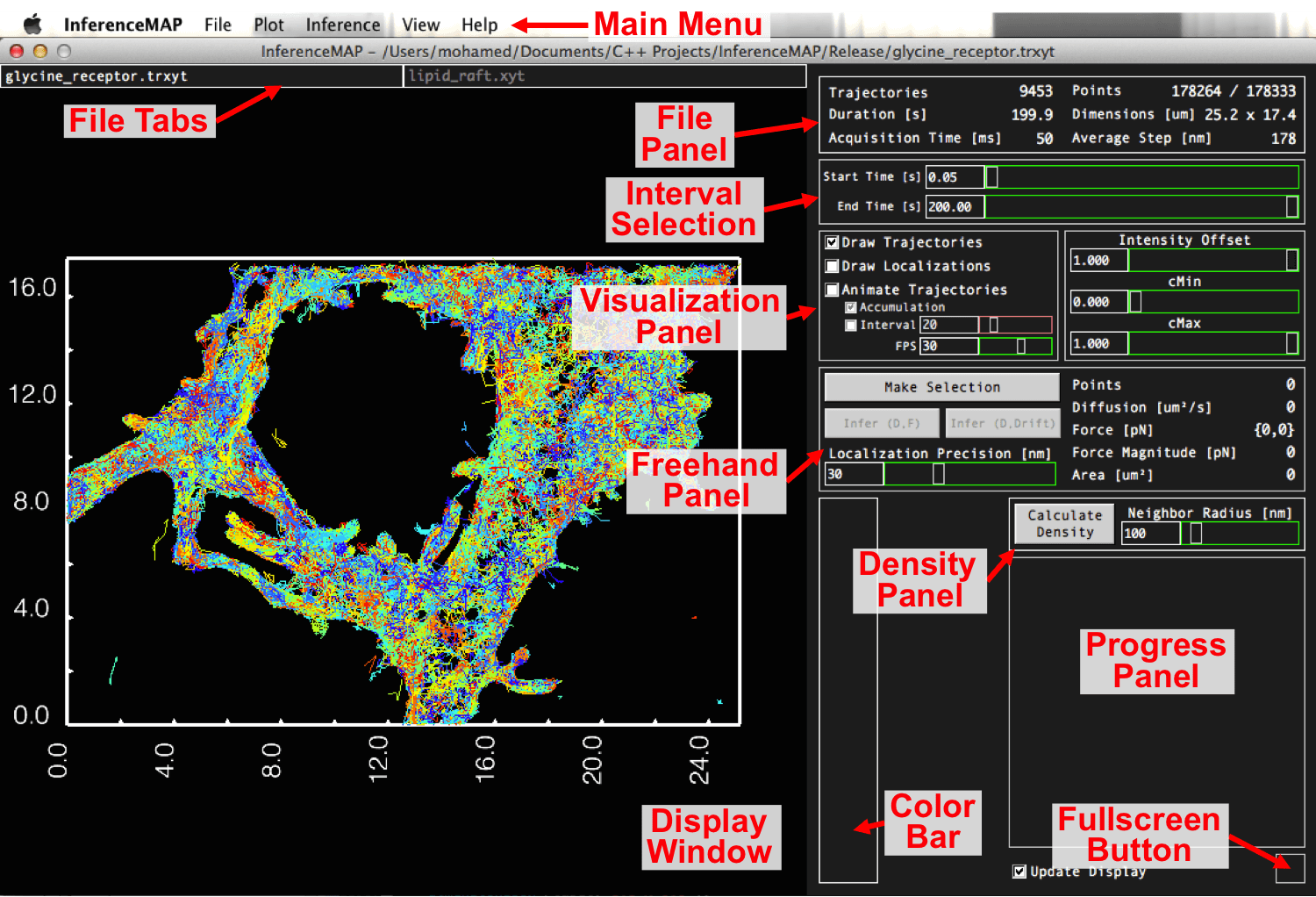}}
\caption{\label{fig:main_interface} \softwareName main interface.}
\end{figure}

\softwareName is controled with an interactive graphical user interface, as shown in Figure \ref{fig:main_interface}.

\begin{description}
	\item \textbf{Main Menu} -- Access to input/output options and all the main functions in \softwareName
	\item \textbf{Display Window} -- Window displaying current trajectory data set
	\item \textbf{File Tabs} -- Indication of files opened in current session
	\item \textbf{File Panel} -- Spatiotemporal data concerning the current file
	\item \textbf{Interval Selection} -- Select to activate trajectories within a specified time window
	\item \textbf{Freehand Panel} -- Macro for selecting a custom region in the current file and performing a direct trajectory inference
	\item \textbf{Color Bar} -- Color code corresponding to the selected meshing parameter overlay
	\item \textbf{Visualization Panel} -- Current file viewing options 
	\item \textbf{Density Panel} -- Calculate the localization density for the displayed trajectories
	\item \textbf{Progress Panel} -- Displays calculation progress
	\item \textbf{Fullscreen Button} -- Toggle between full and default screen size
\end{description}

\clearpage

\section{Trajectory Visualization}
\label{sec:trajectory_visualization}

\begin{figure}[h!]
\center{\includegraphics[scale=0.6]{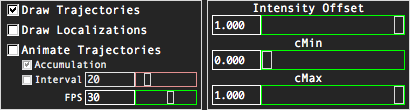}}
\caption{\label{fig:visualization_panel} Visualization Panel in the main interface of \softwareName.}
\end{figure}

Upon starting \textbf{InferenceMAP}, trajectories may be inspected with various visualization functions. After loading a trajectory file (via the \textbf{File} menu), data may be visualized with options in the \textbf{Visualization Panel}, shown in Figure \ref{fig:visualization_panel}.

For a loaded trajectory, the user has access to the following visualization options:
\begin{description}
	\item \textbf{Draw Trajectories.} Displays all trajectories in loaded file. For \textbf{.trxyt} files, trajectories are distinguished by different colors.
\begin{figure}[h!]
\center{\includegraphics[scale=0.22]{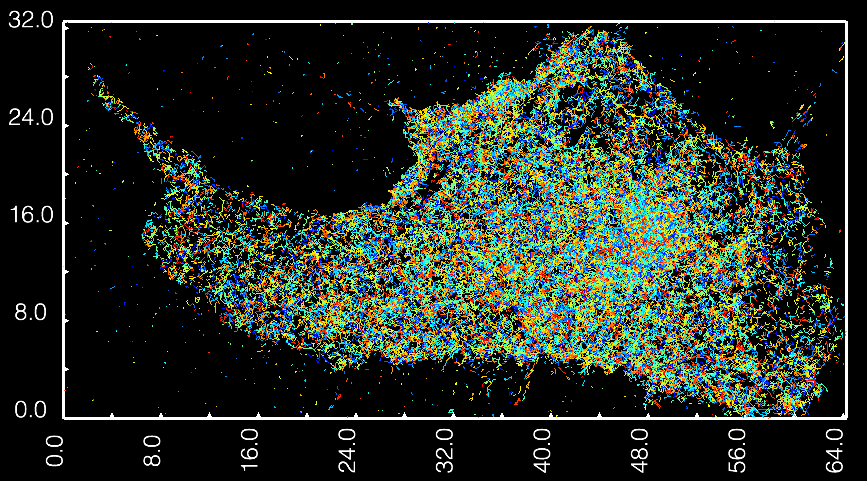}}
\end{figure}
	\item \textbf{Draw Localizations.} Draws localizations (points) from all trajectories (color code is corresponds to time of appearance).
\begin{figure}[h!]
\center{\includegraphics[scale=0.22]{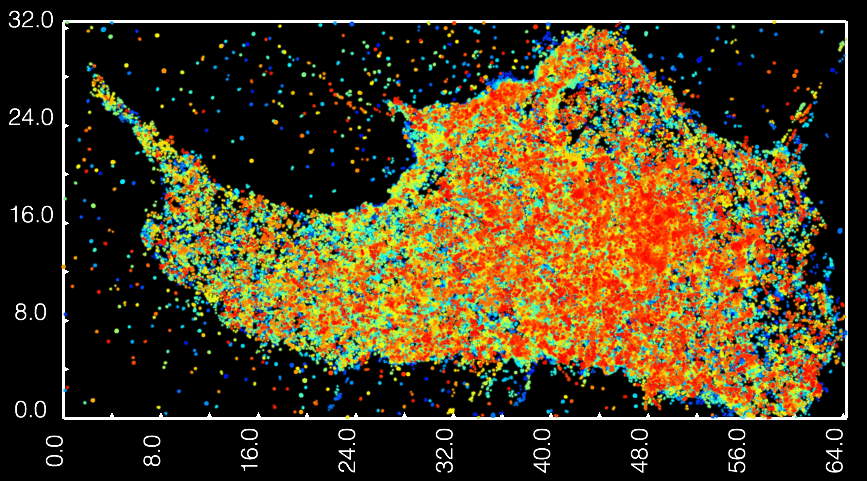}}
\end{figure}
	\item \textbf{Animate Trajectories.} Animates the trajectories either in: \textbf{Accumulation} mode which animates and accumulates all the trajectories in the loaded file, or in an \textbf{Interval} mode which animates the trajectories in a ``sliding window'' mode, in which the interval size may be adjusted with its corresponding slider (20 steps by default). The \textbf{FPS} slider adjusts the frames-per-second of animations in the \textbf{Display Window}.
\begin{figure}[h!]
\center{\includegraphics[scale=0.22]{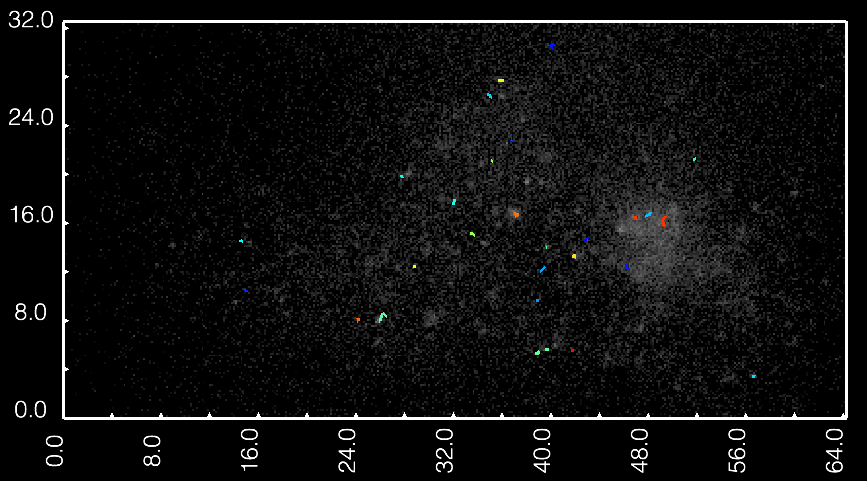}}
\end{figure}
\end{description}

Trajectory viewing may additionally be synchronized to raw acquisition films, using the \textbf{TIFF Overlay} tool described in Section \ref{sec:tiff_overlay}.

\clearpage

\section{Meshing}
\label{sec:meshing}

\begin{figure}[h!]
\center{\includegraphics[scale=0.35]{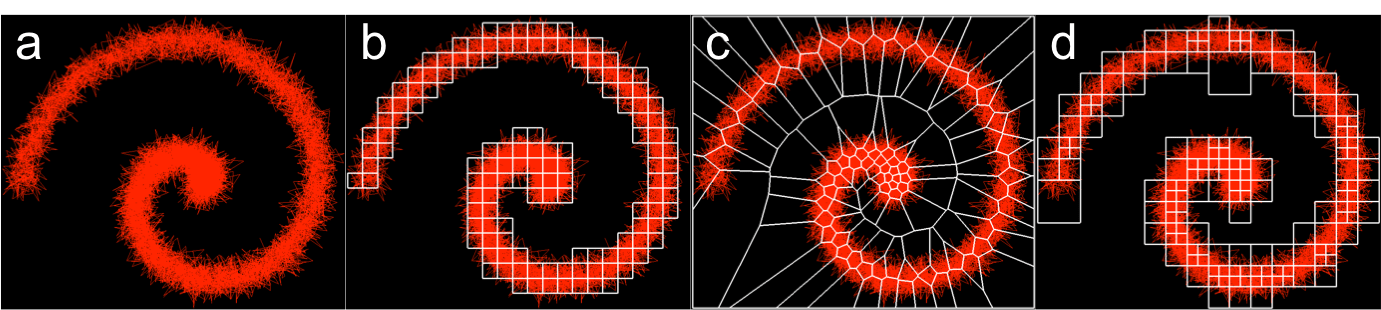}}
\caption{\label{fig:meshing} Meshing types available in \softwareName for a simulated spiral trajectory (a): (b) Square Meshing, (c) Voronoi Meshing, and (d) Quad-Tree Meshing}
\end{figure}

\softwareName offers three types of meshing, shown in Figure \ref{fig:meshing}. The type of meshing method deemed appropriate for a given data set depends principally on the density of localizations (not necessarily trajectories), and, not unrelated, the desired resolution with which the dynamic landscape is mapped. The user is reminded that the main assumption in meshing is that the inferred parameter (diffusion coefficient, force components, drift components, or potential energy) remains constant in each zone.

\subsection{Square Meshing}
\label{sec:square_meshing}

\begin{figure}[h!]
\center{\includegraphics[scale=0.6]{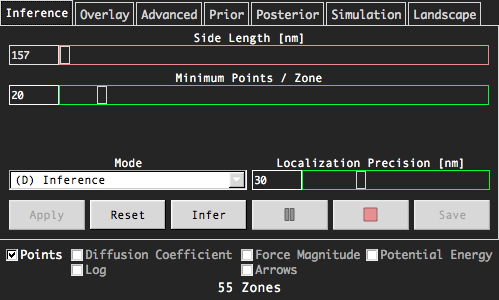}}
\caption{\label{fig:square_meshing} \textbf{Square Meshing} interface.}
\end{figure}

\textbf{Square Meshing} involves spatial partitioning the trajectory overlay space into identically-sized squares. It is most appropriate in cases where the localization density throughout the data set is relatively homogeneous. The \textbf{Square Meshing} dialog box is shown in Figure \ref{fig:square_meshing}.

An appropriate option for the square side length is the average trajectory step length (this is the minimal and initial default value in the \textbf{Side Length} slider of the \textbf{Square Meshing} interface). However, if trajectory points (localizations) are not sufficiently dense, the side length should be increased (at least 20 points per zone is recommended for accurate parameter estimation).\\
\\
To apply a square mesh to the current file, press the \textbf{Apply} button. This will create the internal data structure necessary to perform the inference calculation.\\
\\
The \textbf{Minimum Points / Zone} slider permits filtering of mesh zones based on the number of localizations contained in them. It is generally recommended to have roughly $\sim$ 20 points per cell. This notably adds constraints to the selected zone side length. In the case that the side length needs to be readjusted, the user may select the \textbf{Reset} button, which reinitializes the mesh selection options.\\
\\
The fields in the \textbf{Mode} drop-down menu are described in detail in Section \ref{sec:inference}.\\
\\

\subsection{Voronoi Tessellation}
\label{sec:voronoi_tessellation}

\textbf{Voronoi Tessellation} is one of the adaptive meshing methods available in \textbf{InferenceMAP}. It is most appropriate in cases where there is significant heterogeneity in the density of localizations (not necessarily trajectories). In contrast to \textbf{Square Meshing}, it will generate more zones in regions where localizations are more dense, and additionally adapt the size of zones based on the density of localizations within it. that eventually mesh generation includes two steps: \textbf{Clustering} and \textbf{Tessellation} which are discussed in Sections \ref{sec:clustering} and \ref{sec:tessellation}, respectively.

\begin{figure}[h!]
\center{\includegraphics[scale=0.6]{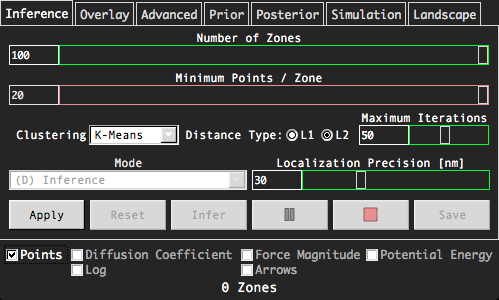}}
\caption{\label{fig:voronoi_tessellation} \textbf{Voronoi Tessellation} interface.}
\end{figure}

The Voronoi tessellated mesh is generated by pressing the \textbf{Apply} button. This will create the internal data structure necessary to perform the inference calculation.

\subsubsection{Clustering}
\label{sec:clustering}

The first step to generate a Voronoi tessellated mesh is to cluster (or group) localizations together in a supervised fashion. The different modes of clustering are available in the \textbf{Clustering} drop-down menu in the \textbf{Voronoi Tessellation} interface (Figure \ref{fig:voronoi_tessellation}). The available modes include:

\begin{itemize}
	\item \textbf{K-Means} -- Localizations are clustered to globally minimize the within-cluster sum of squares of all the clusters.
	\item \textbf{H-Means} -- A simpler algorithm than K-Means, which assigns localizations to the closest randomly-chosen cluster centers. After assignment, clusters centers are replaced by the centroid, and the process is iteratively repeated.
\end{itemize}

Additionally, there is the option to choose between square-distance minimization (L2) or absolute distance minimization (L1), available in the \textbf{Voronoi Tessellation} interface, seen in Figure \ref{fig:voronoi_tessellation}. Maximum clustering iterations can also be specified by the user in the \textbf{Maximum Iterations} slider.\\

The clustering algorithms used are based on those from John Burkardt, available at \url{http://people.sc.fsu.edu/~jburkardt/c_src/asa136/asa136.html}.\\
\\
An additional reference for Voronoi Tessellation is \emph{Spatial Tessellations: Concepts and Applications of Voronoi Diagrams} by Atsuyuki Okabe, et al.

\subsubsection{Tessellation}
\label{sec:tessellation}

The number of clusters (which will correspond to the number of cells in the generated mesh) is specified with the \textbf{Number of Zones} slider in the \textbf{Voronoi Tessellation} interface, seen in Figure \ref{fig:voronoi_tessellation}. Based on the clusters defined from Section \ref{sec:clustering}, a Voronoi diagram is generated. Effectively, it involves spatially partitioning clusters into convex polygons, such that each of the clustered points inside the polygon is closest to its associated barycenter than to any  other.\\
\\
An important point to consider is that there is no restriction on the minimal dimensions of the Voronoi polygons, meaning the characteristic dimension of zones in the voronoi mesh may end up smaller than the average trajectory step-size. If this is observed, the simplest way to avoid the appearance of such zones is to regenerate the mesh with a smaller number of cells (specified in the \textbf{Number of Zones} slider). In practice, however, even with such zones present in the generated mesh, the final calculated parameter values are not greatly perturbed.

\subsection{Quad-Tree Meshing}
\label{sec:Quad-Tree Meshing}

\textbf{Quad-tree Meshing} is an adaptive meshing method that recursively generates subzones (\emph{leaves}) based on a localization \emph{capacity} metric. Being a meshing technique that conforms to the density of localizations, it is especially relevant in cases where trajectory densities are strongly heterogeneous.

\begin{figure}[h!]
\center{\includegraphics[scale=0.6]{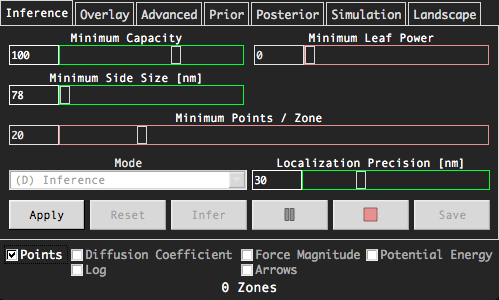}}
\caption{\label{fig:quadtree_meshing} \textbf{Quad-Tree Meshing} interface.}
\end{figure}

Algorithmically, the quad-tree mesh is generated through the addition of localizations into a single square region. Points are added sequentially, until the \emph{capacity} is exceeded. At this point, the mesh is subdivided into four identical squares. This process recursively takes place as more points are added, until no zones exceed the user-specified capacity, although some constraints may prevent this from being the case in practice (see below). The result is a hierarchical mesh.\\

A few constraints (related to trajectory overlay data) distinguish the quad-tree implementation in \softwareName to those that may be used for data structures and other applications:

\begin{itemize}
	\item The side length of zones are limited (by default) to the average trajectory step size of the trajectories. This setting may be adjusted by selecting the minimum side length in the \textbf{Minimum Side Size} slider.
	\item There is an additional iteration at the end of the generation of the quad-tree mesh, which will act to ensure subdivided zones have above the minimal number of points required (by stepping up the quad-tree structure). This is why certain zones may have more than the capacity defined in the \textbf{Minimum Capacity} slider.
	\item Subzones may be filtered based on their \emph{power} (side length) with the \textbf{Minimum Leaf Power} slider.
\end{itemize}

The quad-tree mesh is generated by pressing the \textbf{Apply} button. This will create the internal data structure necessary to perform the inference calculation.\\

\subsection{Meshing Advice}
\label{sec:meshing_advice}

In general, there are some key considerations the user should keep in mind when generating a mesh for the inference calculation described in Section \ref{sec:inference}. This section describes these considerations in some detail.

\subsubsection{Neighboring Zone Connections}
\label{sec:neighboring_zone_connections}
\begin{figure}[h!]
\center{\includegraphics[scale=0.6]{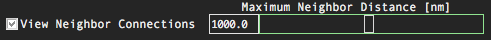}}
\caption{\label{fig:neighbor_connections} Neighbor connection options in \textbf{Advanced} tab of meshing interface.}
\end{figure}

Calculation of the potential (interaction) energy in a given zone necessarily depends on the potential values in neighboring zones. \softwareName enables the user to define the neighbors of given zones via the \textbf{Advanced} tab of the meshing interface window, shown in Figure \ref{fig:neighbor_connections}. Invalid neighbor connections can bias the inference calculation and may reduce the accuracy of the potential energy map.\\

\begin{figure}[h!]
\center{\includegraphics[scale=0.35]{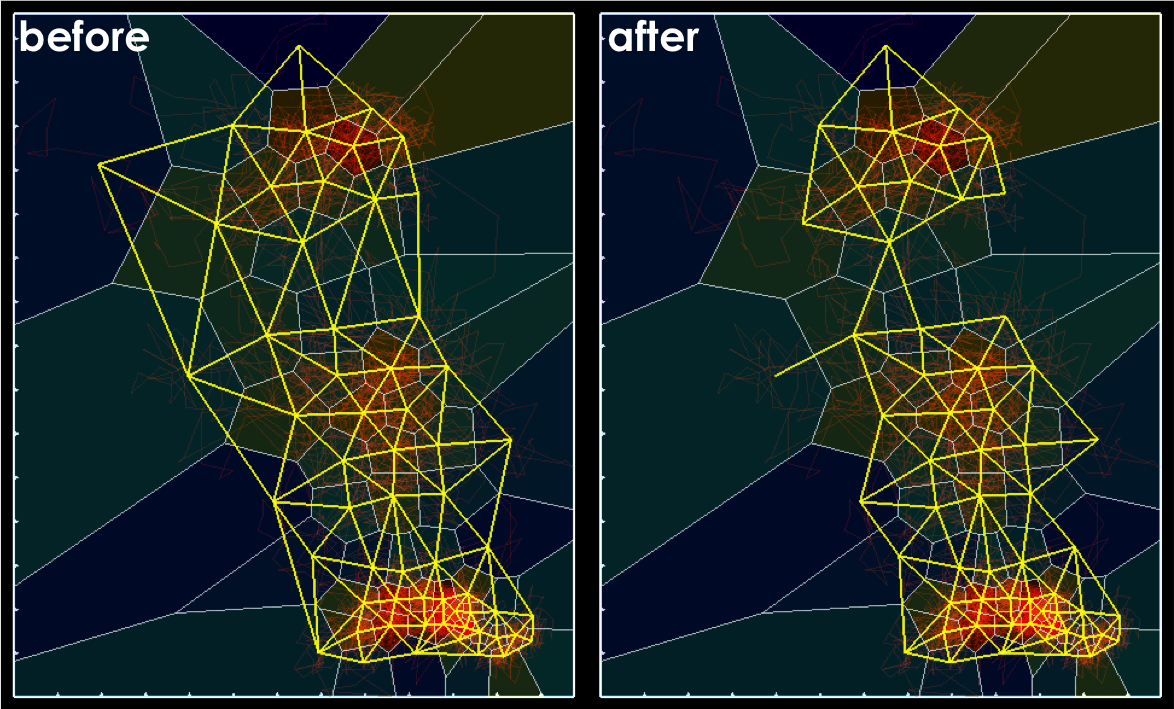}}
\caption{\label{fig:connections} In adjusting the \textbf{Maximum Neighbor Distance} slider, connections between neighboring zones can be added or removed (indicated with yellow lines). This figure demonstrates the removal of erroneous neighboring zone connections.}
\end{figure}

In adjusting the \textbf{Maximum Neighbor Distance} slider, the user can remove connections between zones that are not physically related. These connections are indicated with yellow lines connected between the barycentres of the localizations in each mesh zone.\\
\\
These erroneous connections can be particularly apparent in the adaptive meshing types (the \textbf{Quad-Tree} and \textbf{Voronoi Tessellation}). Figure \ref{fig:connections} shows how erroneous connections may be removed between zones to ensure a more accurate potential calculation.

\subsubsection{Manual Zone Activation and Deactivation}
After a mesh has been applied, prior to the inference calculation, the user may \emph{manually} activate and deactivate zones. The motivation here is to add flexibility to the otherwise non-flexible zone filtering options (e.g. \textbf{Minimum Points}, \textbf{Minimum Leaf Power}, \textbf{Maximum Neighbor Distance}, etc.). Mesh zones are made active or inactive by selection or deselection via \emph{right-clicking} with the mouse.

\subsubsection{Preventing Holes in the Mesh}
In general, ``holes'' (inactive zones surrounded by active ones) should be avoided. This may perturb the potential calculation, as it can greatly affect local values of the potential gradient that may propagate to other zones.\\
\\
Holes may appear if zones do not contain the minimal number of points specified with the \textbf{Minimum Points} slider in the meshing interface, for example. In this case, it is recommended to manually reactivate these zones (by right-clicking on them).\\
\\
An additional cause for the appearance of holes is if meshing parameters are not appropriately selected:

\begin{itemize}
	\item \textbf{Square Meshing} -- The side length is chosen too small.
	\item \textbf{Voronoi Tessellation} -- To many cells are chosen considering the number of localizations in the loaded file.
	\item \textbf{Quad-Tree Meshing} -- The minimum capacity or minimum side length is chosen too small.
\end{itemize}

\begin{figure}[h!]
\center{\includegraphics[scale=0.35]{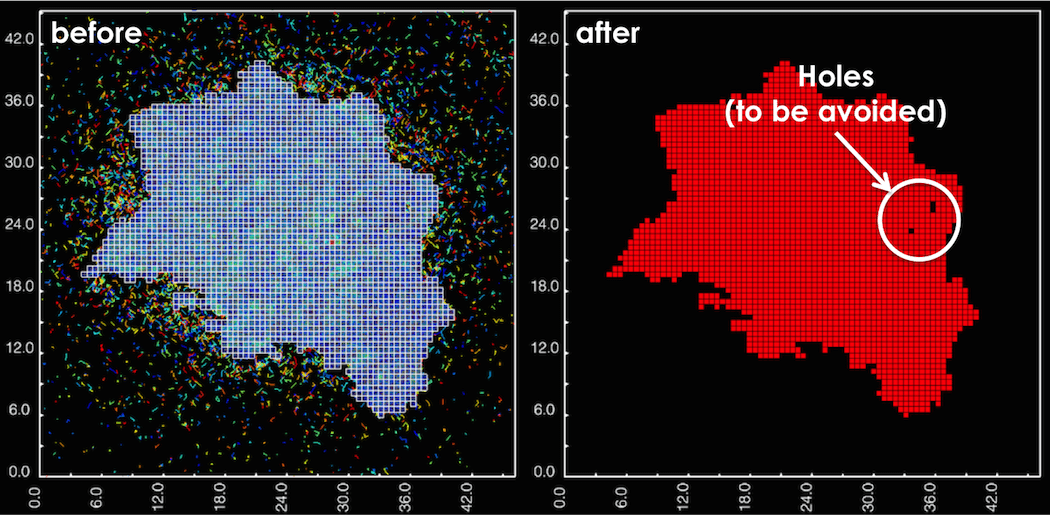}}
\caption{\label{fig:mesh_holes} In adjusting visualization parameters, holes in the mesh can be made obvious. Holes such as these should be avoided (to activate the inactivate zones, it suffices to right click on them).}
\end{figure}

For large maps containing thousands of zones, an easy way to see the inactive zones is by setting the \textbf{cMax} slider to zero, deselecting the \textbf{Draw Trajectories} check button, and changing the grid color or turning it off in the \textbf{Overlay} tab of the meshing interface. Holes in the mesh are made much more evident, as shown in Figure \ref{fig:mesh_holes}.

\clearpage

\section{Inference}
\label{sec:inference}

\softwareName uses the Bayesian inference technique that was first described in \cite{masson2009}. We model the motion of single particles with an overdamped Langevin equation:

\begin{equation}
\label{eqn:overdamped_langevin}
\frac{d\vec{r}}{dt} = \frac{\vec{F}\left( \vec{r} \right)}{\gamma(\vec{r})} + \sqrt{2D(\vec{r})}\xi(t)
\end{equation}

Where $\vec{r}$ is the particle displacement vector, $\vec{F}(\vec{r})$ is the spatially-dependent force (directional bias), $\gamma(\vec{r})$ is the spatially-dependent friction coefficient or viscosity, $D$ is the spatially-dependent diffusion coefficient, and $\xi(t)$ is a zero-average Gaussian noise term. In our case, we may model forces from a potential as $\vec{F}(\vec{r})=-\nabla V$.\\
\\
The associated Fokker-Planck equation, which governs the time-evolution of the particle transition probability $P(\vec{r_{2}},t_{2} | \vec{r_1},t_{1})$, is given by:

\begin{equation}
\label{eqn:fokker_planck}
\frac{dP(\vec{r_{2}},t_{2} | \vec{r_1},t_{1})}{dt}  = -\nabla \cdot \left( -\frac{\nabla V{(\vec{r})}}{\gamma(\vec{r})} P(\vec{r_{2}},t_{2} | \vec{r_1},t_{1}) -\nabla(D(\vec{r})P(\vec{r_{2}},t_{2} | \vec{r_1},t_{1})) \right)
\end{equation}

There is no general analytic solution to Equation \ref{eqn:fokker_planck} for an arbitrary diffusion coefficient $D$ and potential energy $V$. However, if we spatially partition (mesh) the area explored by the single-particle trajectory, we may assume a constant $D$ and $V$ within each partition, upon which the general solution to Equation \ref{eqn:fokker_planck} is a Gaussian, described in:

\begin{equation}
\label{eqn:transition_probability}
P \left( (\vec{r_2},t_{2} | \vec{r_1},t_{1}) | D_{i,j},V_{i,j}\right ) = \frac{\exp -\left( \frac{\left( \vec{r_2}-\vec{r_1}-\frac{\nabla V_{i,j}(t_2-t_1)}{\gamma_{i,j}} \right)^{2}}{4\left( D_{i,j}+\frac{\sigma^2}{(t_2-t_1)}\right )(t_2-t_1) } \right)}{4\pi\left( D_{i,j}+\frac{\sigma^2}{(t_2-t_1)}\right )(t_2-t_1) }
\end{equation}

Where $i,j$ represent indices for the zones of the mesh and $\sigma$ represents the experimental localization precision. An advantage to this approach is that each mesh zone is free to have a different $D$ and $V$ (they are not necessarily constant over the entirety of the trajectory).\\

The overall probability of a trajectory $T$ due to the spatially dependent variables $D_{i,j}$ and $V_{i,j}$ is computed by multiplying the probabilities of all the individual subdomains $P(T|D_{i,j},V_{i,j})$ to give an expression for the likelihood:

\begin{equation}
\label{eqn:likelihood}
P\left( T|D,V \right ) = \prod_{i,j} P\left( T|D_{i,j},V_{i,j}\right)
\end{equation}

With Equation \ref{eqn:likelihood} we apply Bayes' Rule, which states:

\begin{equation}
\label{eqn:bayes_rule}
P \left( D,V|T \right) = \frac{P\left(T|D,V\right)P\left(D,V\right)}{P\left(T\right)}
\end{equation}

Where $P \left( D,V|T \right)$ is the \emph{posterior probability}, $P\left(D,V\right)$ is the \emph{prior probability}, and $P\left(T\right)$ is the \emph{evidence} (which is treated as a normalization constant).\\

For each mesh zone, we perform an optimization of the posterior probability $P \left( D,V|T \right)$ for the model parameters $D$, $V$ (or $\vec{F}$). This is the \emph{maximum a posteriori} (\textbf{MAP}) Bayesian inference approach which is used in \textbf{InferenceMAP}.

\subsection{Inference Workflow}
\label{sec:inference_workflow}

\softwareName offers different modes of performing the inference calculation. Selection between the different modes depends largely on the biological context and what dynamic information the user desires to have extracted. Table \ref{tab:inference_modes} summarizes the features of each of the inference modes.

\begin{center}
\begin{table}[h!]
\centering
\caption{ \label{tab:inference_modes} Summary of different inference modes available in \softwareName.}
\renewcommand{\arraystretch}{2.75}
\begin{tabular}{l|l|l|l|l}
 \textbf{Inference Mode} & \textbf{Parameters} & \TitleParbox{\textbf{Speed}} & \TitleParbox{\textbf{Priors}} & \TitleParboxA{\textbf{Generated Maps}} \\
  \hline
  (D) & D & Fast & \TitleParbox{Uniform, Jeffreys, Smoothing} & \TitleParboxA{Diffusion Maps} \\
  (D,F) & D, F, V & Fast & \TitleParbox{Uniform, Jeffreys, Smoothing} & \TitleParboxA{Diffusion, Force and \emph{indirect} Potential} \\
  (D,Drift) & D, $\frac{\vec{F}}{\gamma}$ & Fast & \TitleParbox{Uniform, Jeffreys, Smoothing} & \TitleParboxA{Diffusion and Drift} \\
  (D,V) & D, F, V & Medium & 	\TitleParbox{Uniform, Jeffreys, Smoothing}	& \TitleParboxA{Diffusion, Conservative Force and Potential} \\
  Polynomial Potential &  D, F, V & Slow & \TitleParbox{Uniform, Jeffreys} & \TitleParboxA{Diffusion, Force and Confined Potential} \\
\end{tabular}
\end{table}
\end{center}

\begin{figure}[h!]
\center{\includegraphics[scale=0.6]{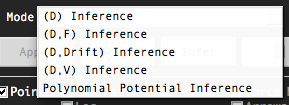}}
\caption{\label{fig:inference_mode} Inference mode selection in meshing interface.}
\end{figure}

The inference mode is selected from the \textbf{Mode} drop-down menu in the chosen meshing interface (Figure \ref{fig:inference_mode}).\\

To perform the inference calculation for a mesh, the following steps are generally taken:

\begin{enumerate}
	\item The mesh is applied by pressing the \textbf{Apply} button of the meshing interface
	\item Adjustments to the mesh may be made as described in Section \ref{sec:meshing_advice} (e.g. activating or deactivating selected regions, choosing neighbour connections, etc.)
	\item The localization precision is specified with the \textbf{Localization Precision} slider in the \textbf{Inference} tab of the meshing interface.
	\item Priors may be selected or deselected in the \textbf{Prior} tab of the meshing interface.
	\item Depending on the number of zones and the inference mode selected, \textbf{Randomized Optimization} may be activated in the \textbf{Advanced} tab of the meshing interface.
	\item Physical parameters are inferred by pressing the \textbf{Infer} button in the \textbf{Inference} tab of the meshing interface.
\end{enumerate}

\subsection{(D) Inference}
\label{sec:d_inference}

The \textbf{(D) Inference} mode estimates solely the diffusion coefficient in the active zones of a mesh. Diffusion is estimated in each zone independently from the others, resulting in a rapid calculation as this mode consists of a single-variable optimization. The posterior probability used to infer the diffusion in a given zone is given by:

\begin{equation}
\label{eqn:d_inference}
P\left(\lbrace{  D_{i,j}  \rbrace}|\lbrace{T_{k}\rbrace} \right) \propto 
\prod_{k} \prod_{\mu:\vec{r}^{k}_{\mu}\in S_{i,j}}  \frac{  \exp\left(-\frac{  \left( \vec{r}^{k}_{\mu + 1} - \vec{r}^{k}_{\mu} \right)^{2}    }{ 4\left(D_{i,j}+\frac{\sigma^{2}}{\Delta t}\right)\Delta t     }      \right)               }{4\pi\left(D_{i,j}+\frac{\sigma^{2}}{\Delta t}\right)\Delta t } \times P_{J}\left( D_{i,j} \right) \times P_{S}\left( D_{i,j} \right)
\end{equation}

Where $D$ is the diffusion coefficient, $\mu$ designates the index for which the points $\vec{r}^{k}_{\mu}$ of the $k^\mathrm{th}$ trajectory are in $S_{i,j}$ (the current zone being analyzed), and $\sigma$ is the experimental localization accuracy (30~nm by default).\\
\\
The $P_{J}\left( D_{i,j} \right)$ term in (\ref{eqn:d_inference}) designates the optional Jeffreys' prior (Section \ref{sec:jeffreys_prior}). Jeffreys' prior may be activated and deactivated in the \textbf{Priors} tab of the meshing interface.\\
\\
The $P_{S}(D_{i,j})$ term is the diffusion smoothing prior which is described in Section \ref{sec:smoothing_prior}.\\
\\
The final result of this calculation is the maximum a posteriori (MAP) estimate, $D^{MAP}_{i,j}$, and is updated to the mesh in the display window.\\

\subsubsection{Applicability}
The \textbf{(D) Inference} mode is well-suited to trajectories in the following situations:
\begin{itemize}
	\item Freely diffusing molecules
	\item Rapid characterization of the diffusivity
\end{itemize}

\subsection{(D,F) Inference}
\label{sec:df_inference}
The \textbf{(D,F) Inference} mode estimates the diffusion coefficient and force components in the active zones of a mesh. Parameters are estimated in each zone independently from the others. Specifically, the Bayesian inference effectuated in each zone \emph{i,j} is an optimization of $D$ and $\vec{F}$ in the zonal posterior probability:

\begin{equation}
\label{eqn:df_inference}
P\left(\lbrace{  D_{i,j}  \rbrace},\lbrace{\vec{F}_{i,j}\rbrace}|\lbrace{T_{k}\rbrace} \right) \propto 
\prod_{k} \prod_{\mu:\vec{r}^{k}_{\mu}\in S_{i,j}}  \frac{  \exp\left(-\frac{  \left( \vec{r}^{k}_{\mu + 1} - \vec{r}^{k}_{\mu} - D_{i,j} \vec{F}_{i,j}\Delta t/k_{B}T \right)^{2}    }{ 4\left(D_{i,j}+\frac{\sigma^{2}}{\Delta t}\right)\Delta t     }      \right)               }{4\pi\left(D_{i,j}+\frac{\sigma^{2}}{\Delta t}\right)\Delta t } \times P_{J}\left( D_{i,j} \right) \times P_{S}\left( D_{i,j} \right)
\end{equation}

Where $D$ is the diffusion coefficient, $\vec{F}$ is the force vector, $\mu$ designates the index for which the points $\vec{r}^{k}_{\mu}$ of the $k^\mathrm{th}$ trajectory are in $S_{i,j}$ (the current zone being analyzed), and $\sigma$ is the experimental localization accuracy (30~nm by default).\\
\\
The $P_{J}\left( D_{i,j} \right)$ term in (\ref{eqn:d_inference}) designates the optional Jeffreys' prior (Section \ref{sec:jeffreys_prior}). Jeffreys' prior may be activated and deactivated in the \textbf{Priors} tab of the meshing interface.\\
\\
The $P_{S}(D_{i,j})$ term is the diffusion smoothing prior which is described in Section \ref{sec:smoothing_prior}.\\
\\
The final result of this calculation are the maximum a posteriori (MAP) estimates, $D^{MAP}_{i,j}$ and $\vec{F}^{MAP}_{i,j}$ and are updated to the mesh in the display window.\\

For further information regarding this calculation, see reference \cite{masson2014}.\\

\subsubsection{Potential Calculation}
\label{sec:df_inference_potential_calculation}

The user has the option of estimating the potentials, $V$, following the $D^{MAP}_{i,j}$ and $\vec{F}^{MAP}_{i,j}$ calculation described in Section \ref{sec:df_inference}. Potential values are estimated with a least-squares minimization between $\vec{F}^{MAP}_{i,j}$ (or $\nabla V^{MAP}_{i,j}$) and the gradient of the theoretical values for the potential under thermal equilibrium conditions, $\nabla V_{i,j}$. A user-defined penalization factor, $\beta$, is introduced to penalize the effect of strong potential gradients. A typically used value for $\beta$ is $\sim$ 2.0 (default). It is emphasized that low values of $\beta$ favor large local variations in the potential field, while high values will act to damp large variations.\\
\\
Specifically, the minimization is performed on zones that have at least one neighbor. The calculation minimizes $xi$ as descibed in:

\begin{equation}
\xi \left(V_{i,j}\right) = \sum_{i,j} \left( \nabla V_{i,j} - \nabla V_{i,j}^{MAP} \right)^{2} + \beta \sum \left( \nabla V_{i,j} \right)^{2}
\end{equation}

\begin{figure}[h!]
\center{\includegraphics[scale=0.6]{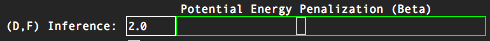}}
\caption{\label{fig:beta} Potential energy penalization factor $\beta$ selection slider.}
\end{figure}

The penalization factor, $\beta$ can be specified in the \textbf{Advanced} tab of the chosen meshing interface (Figure \ref{fig:beta}).

\subsubsection{Applicability}
The \textbf{(D,F) Inference} mode is well-suited to trajectories in the following situations:
\begin{itemize}
	\item Mapping of local force components
	\item The presence of non-potential forces (e.g. a rotational component)
	\item Rapid characterization of the diffusion coefficient and directional biases of the trajectories
\end{itemize}

\subsection{(D,Drift) Inference}
\label{sec:ddrift_inference}

The \textbf{(D,Drift) Inference} mode estimates the diffusion coefficient and drift ($\frac{\vec{F}}{\gamma}$) in the active zones of a mesh. Prameters are estimated in each zone independently from the others. Specifically, the Bayesian inference effectuated in each zone \emph{i,j} is an optimization of $D$ and $\frac{\vec{F}}{\gamma}$ in the zonal posterior probability:

\begin{equation}
\label{eqn:ddrift_inference}
P\left(\lbrace{  D_{i,j}  \rbrace},\lbrace{\frac{\vec{F}_{i,j}}{\gamma}\rbrace}|\lbrace{T_{k}\rbrace} \right) \propto 
\prod_{k} \prod_{\mu:\vec{r}^{k}_{\mu}\in S_{i,j}}  \frac{  \exp\left(-\frac{  \left( \vec{r}^{k}_{\mu + 1} - \vec{r}^{k}_{\mu} - \frac{\vec{F}_{i,j}}{\gamma}\Delta t/k_{B}T \right)^{2}    }{ 4\left(D_{i,j}+\frac{\sigma^{2}}{\Delta t}\right)\Delta t     }      \right)               }{4\pi\left(D_{i,j}+\frac{\sigma^{2}}{\Delta t}\right)\Delta t } \times P_{J}\left( D_{i,j} \right) \times P_{S}\left( D_{i,j} \right)
\end{equation}

Where $D$ is the diffusion coefficient, $\vec{F}$ is the force vector, $\gamma$ is the friction (viscosity), $\mu$ designates the index for which the points $\vec{r}^{k}_{\mu}$ of the $k^\mathrm{th}$ trajectory are in $S_{i,j}$ (the current zone being analyzed), and $\sigma$ is the experimental localization accuracy (30~nm by default).\\
\\
The $P_{J}\left( D_{i,j} \right)$ term in (\ref{eqn:d_inference}) designates the optional Jeffreys' prior (Section \ref{sec:jeffreys_prior}). Jeffreys' prior may be activated and deactivated in the \textbf{Priors} tab of the meshing interface.\\
\\
The $P_{S}(D_{i,j})$ term is the diffusion smoothing prior which is described in Section \ref{sec:smoothing_prior}.\\
\\
The final result of this calculation are the maximum a posteriori (MAP) estimates, $D^{MAP}_{i,j}$ and $\vec{\frac{F}{\gamma}}^{MAP}_{i,j}$ and are updated to the mesh in the display window.\\

\subsubsection{Applicability}
The \textbf{(D,Drift) Inference} mode is well-suited to trajectories in the following situations:
\begin{itemize}
	\item Active processes (e.g. active transport phenomena)
\end{itemize}

\subsection{(D,V) Inference}
\label{sec:dv_inference}

The \textbf{(D,V) Inference} mode directly computes the diffusion coefficient and the gradients of the potential in each active zone in the mesh. Naturally the calculation is expensive as it estimates all the variables (the diffusion coefficients and potential energies) in all the zones of the mesh at the same time. A randomized optimization method (Section \ref{sec:randomized_optimization}) is included to alleviate its generally high computational requirements. The zonal posterior probability is described as:

\begin{equation}
\label{eqn:dv_inference}
P\left(\lbrace{  D_{i,j}  \rbrace},\lbrace{\nabla V_{i,j}\rbrace}|\lbrace{T_{k}\rbrace} \right) \propto 
\prod_{k} \prod_{\mu:\vec{r}^{k}_{\mu}\in S_{i,j}}  \frac{  \exp\left(-\frac{  \left( \vec{r}^{k}_{\mu + 1} - \vec{r}^{k}_{\mu} - D_{i,j} \nabla V_{i,j}\Delta t/k_{B}T \right)^{2}    }{ 4\left(D_{i,j}+\frac{\sigma^{2}}{\Delta t}\right)\Delta t     }      \right)               }{4\pi\left(D_{i,j}+\frac{\sigma^{2}}{\Delta t}\right)\Delta t } \times P_{J}\left( D_{i,j} \right) \times P_{S}\left( D_{i,j},V_{i,j} \right)
\end{equation}

Where $D$ is the diffusion coefficient, $V$ is the potential energy, $\mu$ designates the index for which the points $\vec{r}^{k}_{\mu}$ of the $k^\mathrm{th}$ trajectory are in $S_{i,j}$ (the current zone being analyzed), and $\sigma$ is the experimental localization accuracy (30~nm by default).\\
\\
The $P_{J}\left( D_{i,j} \right)$ term in (\ref{eqn:d_inference}) designates the optional Jeffreys' prior (Section \ref{sec:jeffreys_prior}). Jeffreys' prior may be activated and deactivated in the \textbf{Priors} tab of the meshing interface.\\
\\
The $P_{S}(D_{i,j},V_{i,j})$ term is the diffusion and potential smoothing prior which is described in Section \ref{sec:smoothing_prior}.

\subsubsection{Applicability}
The \textbf{(D,V) Inference} mode is well-suited to trajectories in the following situations:
\begin{itemize}
	\item Systems with stable interaction sites
\end{itemize}

\subsection{Polynomial Potential Inference}
\label{sec:polynomial_inference}

In contexts where there is clear confinement of a tracked particle, it is useful to describe the confining energy as a polynomial of order $C$ in:

\begin{equation}
\label{eqn:polynomial_potential}
V \left( \vec{r} \right) = \sum^{C}_{j=0} \sum^{j}_{i=0} \alpha_{i,j} x^{i} y^{j-i}
\end{equation}

Where the constants $\alpha_{i,j}$ are fitted to the experimental force fields using standard simplex methods.\\
\\
This description of the potential is integrated into the inference calculation as in Equation \ref{eqn:df_inference}. This type of inference is prohibitively expensive for large maps, it is recommended for localized trajectories in small regions. Further information regarding this type of modelling is described in reference \cite{masson2009}.

\subsubsection{Applicability}
The \textbf{Polynomial Potential Inference} mode is well-suited to trajectories in the following situations:
\begin{itemize}
	\item Confined trajectories
\end{itemize}

\subsection{Randomized Optimization}
\label{sec:randomized_optimization}

To tackle large problems consisting of several hundred or thousand of individual zones, a \textbf{Randomized Optimization} function is available, which greatly reduces computation time. It can be used in the following inference modes:

\begin{itemize}
	\item \textbf{(D) Inference} when a smoothing prior is used
	\item \textbf{(D,F) Inference} when a smoothing prior is used
	\item \textbf{(D,Drift) Inference} when a smoothing prior is used
	\item \textbf{(D,V) Inference}
\end{itemize}

This feature is available for all meshing modes, parameters of which can be adjusted in the \textbf{Advanced} tab of the meshing interface, seen in Figure \ref{sec:randomized_optimization}.

\begin{figure}[h!]
\center{\includegraphics[scale=0.6]{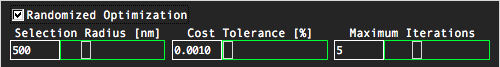}}
\caption{\label{fig:randomized_optimization} \textbf{Randomized Optimization} options in the meshing interface.}
\end{figure}

\begin{figure}[h!]
\center{\includegraphics[scale=0.6]{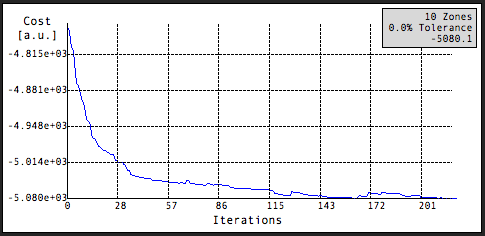}}
\caption{\label{fig:randomized_optimization_cost} \textbf{Randomized Optimization} interface, demonstrating the decay in the cost function. Generally, the \textbf{Randomized Optimization} calculation may be stopped when the cost function has sufficiently decayed, as indicated.}
\end{figure}

To activate the \textbf{Randomized Optimization} mode, a compatible mode must be selected in the \textbf{Inference} tab of the meshing interface (the smoothing prior also needs to be activated for the cases of \textbf{(D)}, \textbf{(D,F)}, and \textbf{(D,Drift)} inference modes). Activation consists in clicking the \textbf{Randomized Optimization} check box seen in Figure \ref{sec:randomized_optimization}.\\
\\
This function works as follows:

\begin{itemize}
	\item Subregions of zones are selected in a circle of radius defined by the \textbf{Selection Radius [nm]} slider (Figure \ref{sec:randomized_optimization}). The zone in which the circle is centered is selected randomly among the activated zones of the mesh. It is advisable to select a radius such that roughly $\sim$10 zones or more will be encompassed.
	\item The inference calculation is performed \emph{only} for the subregion zones, parameters (e.g. diffusion) in all other zones remaining constant. Calculations are limited to the number of iterations specified in the \textbf{Maximum Iterations} slider (Figure \ref{sec:randomized_optimization}). It is recommended to keep the maximum number of iterations between 5--10. Parameter values for the subregion zones are updated after the maximum number of iterations has been reached, thereafter another selection circle is chosen.
	\item A stopping condition may be specified based on the percent error between consecutive posteriori cost function values (defined in the \textbf{Cost Tolerance [\%]} slider). It is recommended to keep this tolerance below 1~\% as the stopping condition may be met before all zones in the mesh have been visited. Often, it is suggested to define a 0~\% tolerance, and to manually stop the calculation by pressing the \textbf{Pause} button in the \textbf{Inference} tab when the cost function has decayed sufficiently, as in Figure \ref{fig:randomized_optimization_cost}. For especially large meshes, stopping the calculation will be delayed -- in these cases the user is advised to press the \textbf{Pause} button, keep the cursor hovering above, and wait!
\end{itemize}

\subsection{Freehand Selection Inference}
\label{sec:Freehand_selection_inference}

\begin{figure}[h!]
\center{\includegraphics[scale=0.35]{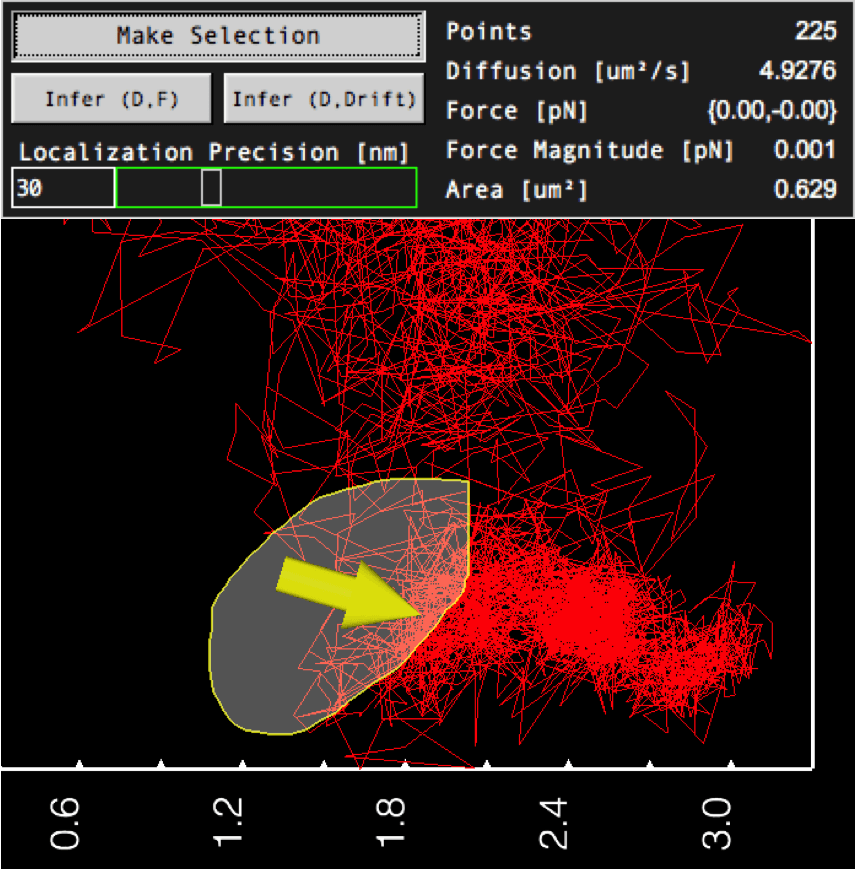}}
\caption{\label{fig:make_selection} \textbf{Freehand Panel} (top) with selected and inferred region (bottom).}
\end{figure}

For determination of the diffusion coefficient and force components in a localized region of a trajectory overlay, \softwareName includes a \textbf{Freehand Selection} tool in the main interface, shown in Figure \ref{fig:make_selection}. This is useful in cases where only subregions of the trajectory space need to be analyzed.\\

To select a freehand region, the \textbf{Make Selection} button is pressed, whereafter the user can select (in lasso style) a closed region of the trajectory overlay by pressing and holding the left mouse button. Once a region has been circled, the user releases the left mouse button (which will close the selected region). In the \textbf{Freehand Panel} (Figure \ref{fig:make_selection}) the user may specify the \textbf{Localization Precision}. Afterwards, the diffusion coefficient and force or drift components determined and displayed by pressing the \textbf{Infer (D,F)} or \textbf{Infer (D,Drift)} buttons, respectively. The inferred region is highlighted in white, with an overlaid arrow to represent the angle of the force or drift (the size of the arrow does not correspond to the magnitude of the force or drift).

\clearpage

\section{Priors}
\label{sec:priors}

\begin{figure}[h!]
\center{\includegraphics[scale=0.6]{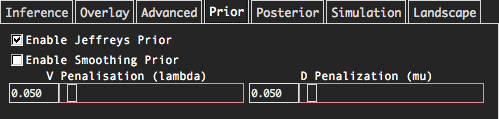}}
\caption{\label{fig:prior_selection} \textbf{Prior} tab of the meshing interface.}
\end{figure}

In general, prior probabilities are used to impose our beliefs to the dynamic parameter estimates that \softwareName performs when generating maps. This section describes three types of prior probabilities that users may incorporate in the inference calculation. They are enabled via the \textbf{Priors} tab of the meshing interfaces, shown in Figure \ref{fig:prior_selection}.

\subsection{Uniform}
\label{sec:uniform_prior}

If no information is known regarding the parameters being inferred (e.g. the diffusivity), the Uniform Prior should be used. It is applied by default if no other priors are activated. Specifically, use of this prior means that the results yielded from the Bayesian inference calculation are identical to those that would be obtained from a maximum likelihood estimation.

\subsection{Jeffreys}
\label{sec:jeffreys_prior}

Jeffreys prior is used to ensure that the posterior probability distribution of an inference calculation is invariant by re-parameterization. Moreover, it allows ``protecting'' inference of the diffusion in cases of high local confinement. In such situations, the ``effective diffusion'' introduced by a non-zero positioning noise can lead to inference of a negative diffusion value \cite{masson2014}. We emphasize that without positioning noise the inference never leads to negative diffusion. Table \ref{tab:jeffreys_priors} lists forms of Jeffreys prior for the different inference modes.\\
\\ 
Jeffreys prior is applicable to a large number of situations. Situations where it is not recommended are in cases where the diffusion is similar to the effective ``noise diffusion'' which we estimate as $\frac{\sigma^{2}}{\Delta t}$. It may be activated and deactivated in the \textbf{Prior} tab of the meshing interface.

\begin{center}
\begin{table}[h!]
\caption{ \label{tab:jeffreys_priors} Forms of Jeffreys' prior for the different inference modes.}
\renewcommand{\arraystretch}{2}
\centering
\begin{tabular}{l|l}
 \textbf{Inference Mode} & \textbf{Jeffreys Prior, $P_{J}$} \\
  \hline
  (D) & $\frac{1}{\left( D_{i,j}\Delta t + \sigma^2 \right)}$ \\
  (D,F) & $\frac{D^{2}_{i,j}}{\left( D_{i,j}\Delta t + \sigma^2 \right)^{2}}$ \\
  (D,Drift) & $\frac{1}{\left( D_{i,j}\Delta t + \sigma^2 \right)^{2}}$ \\
  (D,V) & $\frac{D^{2}_{i,j}}{\left( D_{i,j}\Delta t + \sigma^2 \right)^{2}}$ \\
  Polynomial Potential & $\frac{D^{2}_{i,j}}{\left( D_{i,j}\Delta t + \sigma^2 \right)^{2}}$ \\
\end{tabular}
\end{table}
\end{center}

An important remark is that in the cases of \textbf{D,F} and \textbf{D,V} inference, Jeffreys' prior prevents inferrring negative diffusion coefficients, even when position noise is high.

\subsection{Smoothing}
\label{sec:smoothing_prior}

The smoothing prior penalizes gradients of the physically inferred parameters (either the diffusion or the potential energy, depending on the inference mode). This prior is appropriate for use in biological systems where notions of the strength of diffusion coefficient and potential energy gradients exist. It is meant to reinforce the physical behavior that is to be expected in certain biological systems. For example, in certain situations we do not expect large jumps zone-to-zone in the diffusion coefficient.\\
\\
The diffusion smoothing prior is defined by the surface integral in Equation \ref{eqn:diffusion_smoothing_prior}. The coefficient $\mu$ is the diffusion gradient penalization factor which modulates the strength of smoothing in the area $S$.

\begin{equation}
\label{eqn:diffusion_smoothing_prior}
P_{S}\left(D_{i,j}\right) = \exp \left( -\mu \iint\limits_s ||\nabla D_{i,j}||^{2}ds \right)
\end{equation}

The potential smoothing prior is defined by the surface integral in Equation \ref{eqn:potential_smoothing_prior}. The coefficient $\lambda$ is the potential gradient penalization factor which modulates the strength of smoothing in the area $S$.

\begin{equation}
\label{eqn:potential_smoothing_prior}
P_{S}\left(V_{i,j}\right) = \exp \left( -\lambda \iint\limits_s ||\nabla V_{i,j}||^{2}ds \right)
\end{equation}

Values for $\mu$ and $\lambda$ are specified in the \textbf{Prior} tab of the meshing interface as in Figure \ref{fig:prior_selection}. Table \ref{tab:smoothing_priors} lists the types of smoothing priors available for the different inference modes.

\begin{center}
\begin{table}[h!]
\caption{ \label{tab:smoothing_priors} Forms of smoothing priors for the different inference modes.}
\renewcommand{\arraystretch}{1.5}
\centering
\begin{tabular}{l|l}
 \textbf{Inference Mode} & \textbf{Smoothing Prior, $P_{S}$} \\
  \hline
  (D) & Diffusion \\
  (D,F) & Diffusion \\
  (D,Drift) & Diffusion \\
  (D,V) & Diffusion, Potential \\
  Polynomial Potential & N/A \\
\end{tabular}
\end{table}
\end{center}

Although the \textbf{(D)}, \textbf{(D,F)}, are \textbf{(D,Drift)} modes are generally rapid calculations, the computation time increases substantially when the diffusion smoothing prior is activated. The reason for this is that the problem becomes a calculation in which parameters in all zones (e.~g. diffusion, force, drift) are being optimized simultaneously (without the smoothing prior parameters are optimized in each zone independently which greatly reduces the dimensionality of the problem). For this reason, it may be of interest to use a \textbf{Randomized Optimization} (Section \ref{sec:randomized_optimization}) in large mapping problems which utilize smoothing priors.

\clearpage


\section{Posterior Sampling}
\label{sec:posterior_sampling}

As explained in Section \ref{sec:inference}, the maximum value of the \emph{posterior probability} distribution (the MAP) is used to estimate the dynamic parameters (diffusion coefficient, force components, drift components, and potential energy). Sampling of the \emph{posterior probability} (see Section \ref{sec:inference}) gives insight into the precision of the estimations. Typically, the \emph{posterior probability} distribution takes the approximate form of a Gaussian, where the full-width at half-maximum may be calculated to measure variance of the estimation.\\

\softwareName offers the possibility to sample posterior values for the different parameters. For each of the different meshing types, \emph{after} the inference calculation, the user has the option to sample to posterior probability for the different parameters. Posteriors are sampled for different zones, enabling the user to compare the parameter estimation precision in different parts of the mesh. The different inference modes impose constraints on which parameters may be sampled, as Table \ref{tab:inference_mode_posterior} shows.

\begin{center}
\begin{table}[h!]
\caption{ \label{tab:inference_mode_posterior} Availability of posterior probability sampling for different parameters for different inference modes.}
\begin{tabular}{c|c|c|c|c}
 \textbf{Inference Mode} & \textbf{Diffusion Coefficient} & \textbf{Force Magnitude} & \textbf{Drift Magnitude} & \textbf{Potential Energy} \\
  \hline
  \textbf{(D)} & Yes & No & No & No \\
  \textbf{(D,F)} & Yes & Yes & No & No \\
  \textbf{(D,Drift)} & Yes & No & Yes & No \\
  \textbf{(D,V)} & Yes & No & No & Yes \\
  \textbf{Polynomial Potential} & N/A & N/A & N/A & N/A \\
\end{tabular}
\end{table}
\end{center}

After the inference calculation, posterior probabilities are sampled in the \textbf{Posterior} tab of the meshing interface, shown in Figure \ref{fig:posterior_tab}. The steps to sampling the posterior are the following:

\begin{figure}[h!]
\center{\includegraphics[scale=0.6]{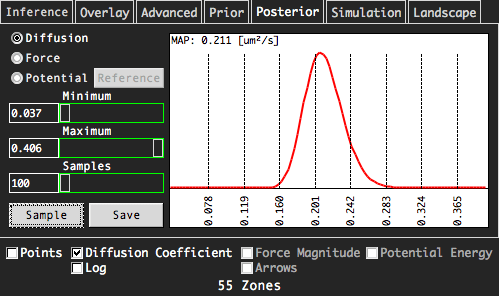}}
\caption{\label{fig:posterior_tab} \textbf{Posterior} tab in the meshing interface.}
\end{figure}

\begin{enumerate}
	\item A zone in the mesh in which to sample the posterior is selected with the mouse in the \textbf{Display Window}
	\item The estimation parameter (diffusion coefficient, force magnitude, or potential energy) is selected in the top-left of the tab
	\item Sampling ranges are selected via the \textbf{Minimum} and \textbf{Maximum} sliders
	\item The number of (equally-spaced) samples are selected with the \textbf{Samples} slider
	\item The posterior for the chosen parameter in the selected zone is sampled by pressing the \textbf{Sample} button, a trace of which is displayed in the interface
	\item Posterior sampling data may be exported to a simple ASCII file by pressing the \textbf{Save} button
\end{enumerate}

For the \textbf{(D,V) Inference} mode, the posterior of the potential may be sampled in two ways:

\begin{enumerate}
	\item A single-zone of the mesh can have its potential sampled across a user-defined range
	\item The posterior of the difference of potentials between two zones may be sampled
\end{enumerate}

\begin{figure}[h!]
\center{\includegraphics[scale=0.35]{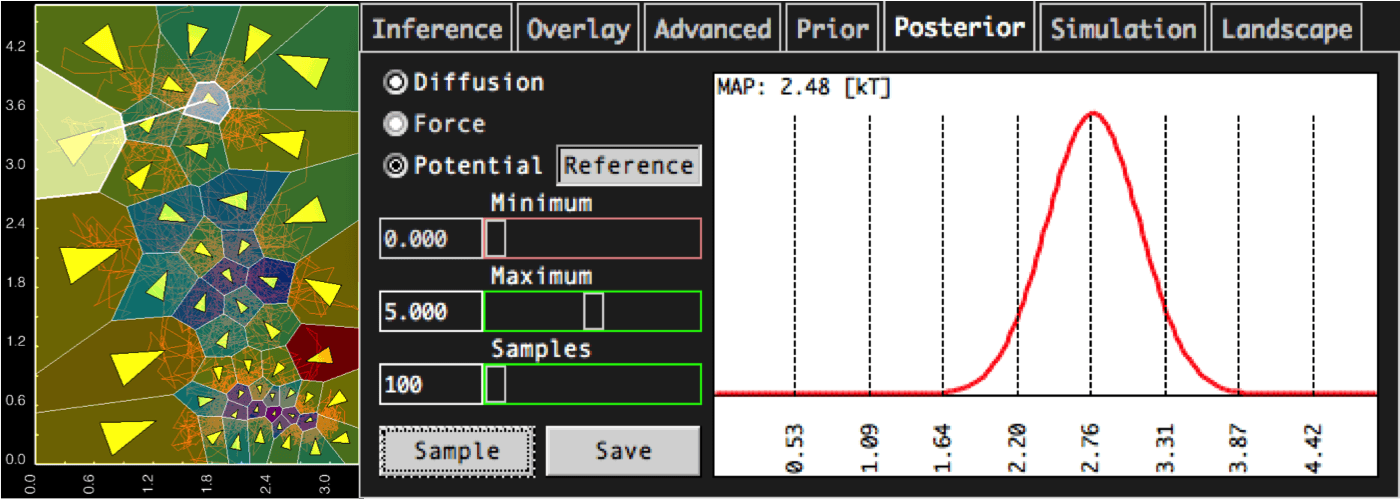}}
\caption{\label{fig:posterior_potential_reference} Posterior sampling of difference of potentials in two zones.}
\end{figure}

For sampling the posterior of the difference of potentials, the first zone is selected with the mouse in the \textbf{Display Window}. Following, by pressing the \textbf{Reference} button, the second may be selected (see Figure \ref{fig:posterior_potential_reference}). At this point, the user may select the range of potential differences to sample between the two zones (specified with the \textbf{Maximum} slider). This type of calculation is useful for measuring the precision of potential energy barriers, for example.

\clearpage

\section{Trajectory Simulation}
\label{sec:inference_trajectory_simulation}

After an inference calculation has been performed to generate diffusion and potential energy maps, there is the possibility to generate simulated trajectories based on the inferred parameter values. This is based on the Gillespie method, described in \cite{masson2014}.\\

The user may be motivated to create simulated trajectories in a few different situations. Namely, in cases where trajectories are short (e.g. less than 10 points) and certain ``long trajectory'' metrics would like to be estimated. These may include measuring whether a trajectory experiences anomalous motion, residence time, and binding and dissociation rates.

\begin{figure}[h!]
\center{\includegraphics[scale=0.6]{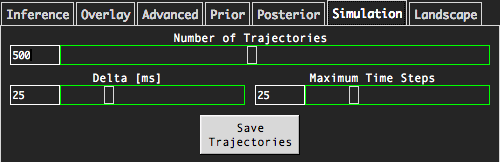}}
\caption{\label{fig:simulation_tab} \textbf{Simulation} tab of the meshing interface.}
\end{figure}

Simulated trajectories are generated via the \textbf{Simulation} tab of the meshing interface, seen in Figure \ref{fig:simulation_tab}.

In this tab, the user may select the number of trajectories, the time spacing (delta) between consecutive points, and the maximum time steps in an individual trajectories. Pressing the \textbf{Save Trajectories} button outputs all trajectories into a \textbf{.trxyt} file (described in Section \ref{sec:file_formats}). Output trajectories will have a spatial resolution corresponding to that of the mesh. \\

\clearpage

\section{Landscape Viewing}
\label{sec:inference_landscape_viewing}

\begin{figure}[h!]
\center{\includegraphics[scale=0.6]{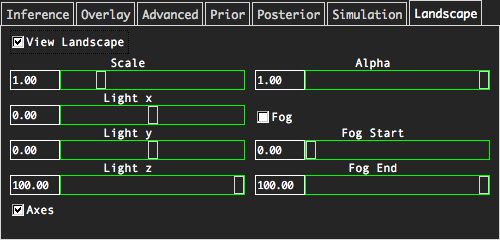}}
\caption{\label{fig:landscape_tab} \textbf{Landscape} tab of the meshing interface.}
\end{figure}

\softwareName offers to option to view dynamical maps as three-dimensional landscapes for more intuitive interpretation. This feature is accessed with the \textbf{Landscape} tab of the meshing interface, shown in Figure \ref{fig:landscape_tab}.\\

\begin{figure}[h!]
\center{\includegraphics[scale=0.35]{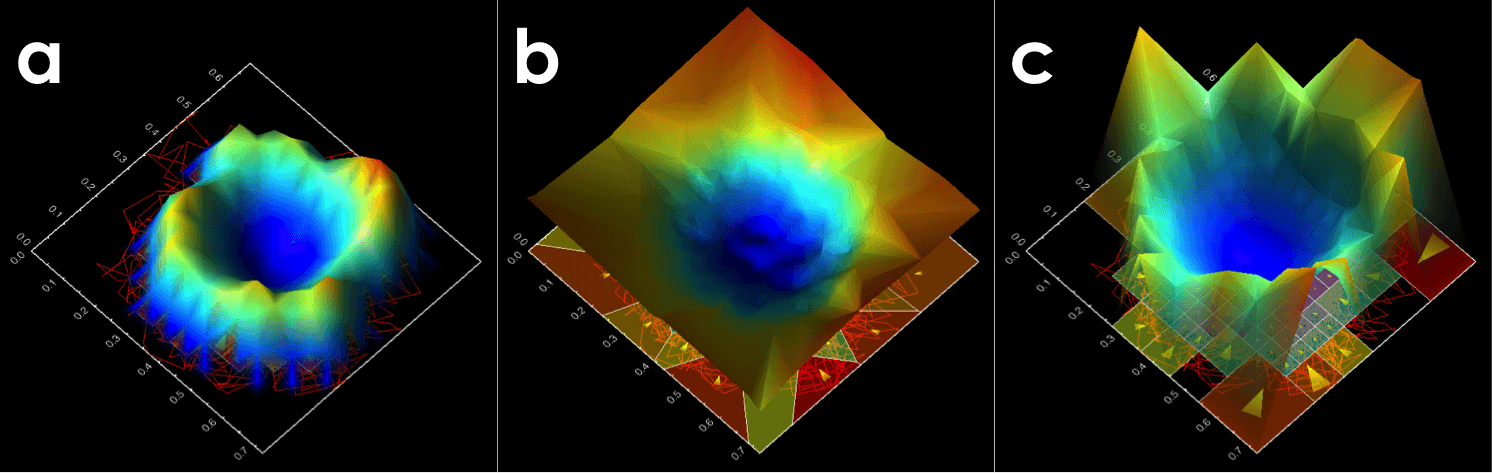}}
\caption{\label{fig:landscape_meshes} Potential energy landscapes of the same trajectory file (a simulated potential well) for the different meshing types: (a) square mesh, (b) Voronoi tessellated mesh, and (c) a quad-tree mesh.}
\end{figure}

Here, the user has various visualization options for viewing the three-dimensional landscape. To change between landscapes, simply press the parameter boxes on the button of the meshing interface. The landscape viewing mode is available for all meshing types (see Section \ref{sec:meshing}). Color scales can be adjusted with the \textbf{cMin} and \textbf{cMax} sliders in the \textbf{Main Interface}. Figure \ref{fig:landscape_meshes} shows the three-dimensional landscapes for the same trajectory file (a simulated potential well).

\clearpage

\section{Freehand Selection}
\label{sec:custom_selection}

\begin{figure}[h!]
\center{\includegraphics[scale=0.6]{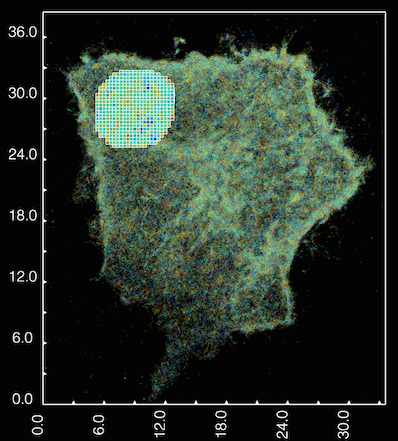}}
\caption{\label{fig:custom_region} Example of a map calculated for a custom-selected subregion.}
\end{figure}

Often it is desired to only generate a dynamical map for a subregion of the entire trajectory space. \softwareName enables users to select custom subregions with the freehand selection tool, by selecting the \textbf{Make Selection} button in the \textbf{Freehand Panel} of the main interface window (top panel of Figure \ref{fig:make_selection}), whereafter the user selects the desired subregion in the \textbf{Display Window} by clicking and holding the right mouse button. Once a closed region has been selected, releasing the right mouse button will create the subregion. At this point, the user opens a meshing interface as usual, by choosing from the options in the \textbf{Inference > Meshing} menu of the \textbf{Main Menu}. The mesh generated in the subregion upon pressing the \textbf{Apply} button in the respective meshing interface. Figure \ref{fig:custom_region} shows an example of a map for a custom-selected region using this feature.\\

It is important to assure that the \textbf{Make Selection} button is toggled \emph{on} before pressing \textbf{Apply} in the chosen meshing interface.


\clearpage

\section{Tools}
\label{sec:tools}

\subsection{Annotations}
\label{sec:annotations}

\begin{figure}[h!]
\center{\includegraphics[scale=0.6]{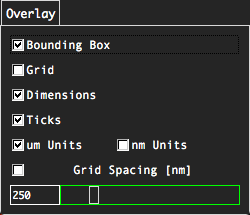}}
\caption{\label{fig:annotations} \textbf{Annotations} interface.}
\end{figure}

Various annotations may be adjusted or removed in the \textbf{Display Window} by accessing the \textbf{Annotations} interface in the \textbf{Tools} menu of the \textbf{Main Menu}, seen in Figure \ref{fig:annotations}.

\subsection{Density Calculation}
\label{sec:density_calculation}
\begin{figure}[h!]
\center{\includegraphics[scale=0.35]{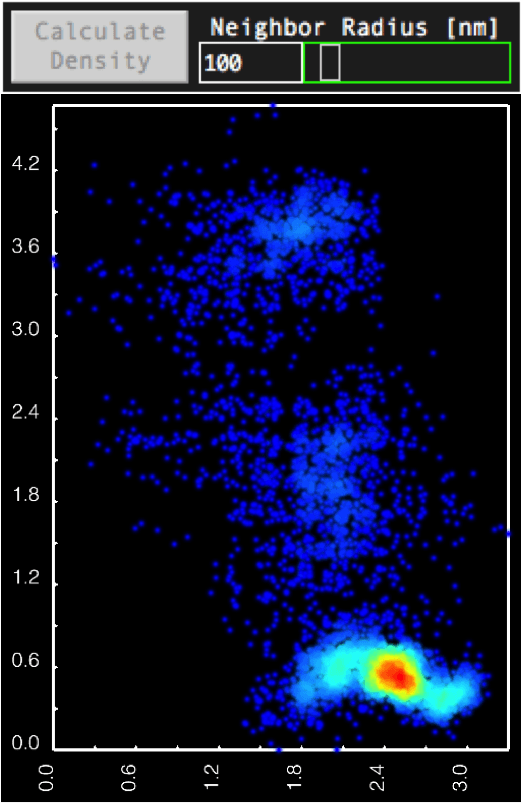}}
\caption{\label{fig:density_calculation} \textbf{Density Panel} (top) in the \textbf{Main Interface}, with localizations density plot (below).}
\end{figure}

The \emph{relative} density of localizations can be calculated by pressing the \textbf{Calculate Density} button in the \textbf{Density Panel} of the \textbf{Main Interface}. The density is determined by counting the number of localizations (neighbors) within the circle defined at each given localization. The size of this circle is determined by the radius indicated in the \textbf{Neighborhood Radius Slider} \textbf{Density Panel} (Figure \ref{fig:density_calculation}).

\subsection{Interval Selection}
\label{sec:interval_selection}
\begin{figure}[h!]
\center{\includegraphics[scale=0.6]{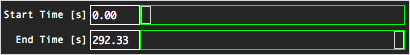}}
\caption{\label{fig:interval_selection} Interval selection sliders in the \textbf{Main Interface}.}
\end{figure}

Specific time intervals in a loaded trajectory dataset can be selected for analysis. This is useful if a temporally windowed analysis is desired. The \textbf{Start Time Slider} and \textbf{End Time Slider} in the \textbf{Main Interface} are used to select the desired interaval. Only trajectories in the selected window will be updated to the display, as shown in Figure \ref{fig:interval_selection}. Afterwards, a mesh may be applied as usual (Section \ref{sec:meshing}), but only trajectories within the selected interval will be considered. Of note, the interval selection is disabled once a mesh has been applied.

\subsection{Save Screen}
\label{sec:save_screen}

At any moment while using \softwareName, a screen shot of the \textbf{Display Window} may be taken by selecting \textbf{Save Screen} in the \textbf{Tools} menu of the \textbf{Main Menu}. This feature will also save a screen shot of the colorbar in a separate file (with \textbf{\_colorbar} postfix).

\subsection{TIFF Overlay}
\label{sec:tiff_overlay}

\begin{figure}[h!]
\center{\includegraphics[scale=0.35]{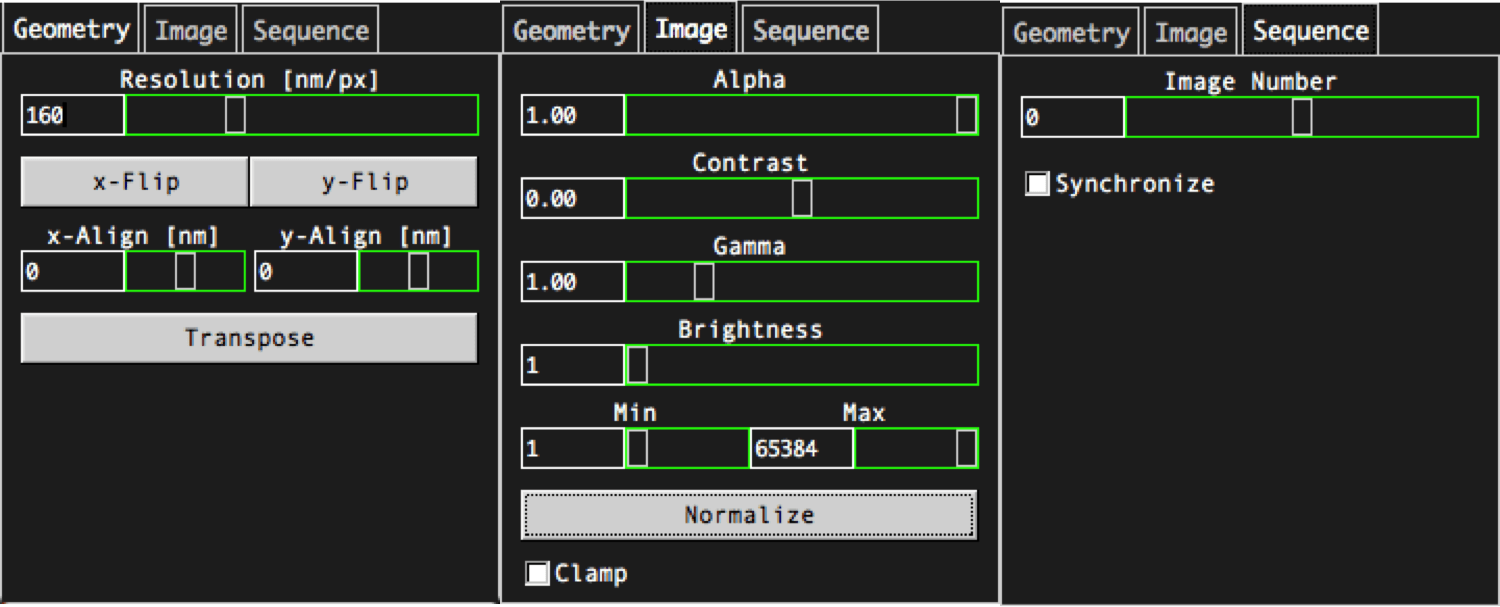}}
\caption{\label{fig:tiff_overlay} \textbf{TIFF Overlay} interface tabs.}
\end{figure}

An important feature in \softwareName is the ability to overlay experimentally-acquired TIFF images to corresponding trajectories. Both single-image files (e.g. DIC or transmission images) and multi-image stacks (such as the raw images from which the trajectories are constructed) may be overlaid to open trajectory files (described in Section \ref{sec:file_formats}). A TIFF image can be overlaid to an open trajectory file, by selecting the \textbf{TIFF Overlay} button in the \textbf{Tools} menu in the \textbf{Main Menu}, shown in Figure \ref{fig:tiff_overlay}. Of note, TIFF images must be 16--bit black and white to be properly overlaid.

The \textbf{Geometry} tab permits adjustment of the pixel size, and allows adjustments to image orientation.\\

The \textbf{Image} tab contains functions for image rendering. The \textbf{Clamp} box clamps the images to the maximal dimensions of the loaded trajectory file.\\

The \textbf{Sequence} tab applies to overlaid multi-image stacks. The \textbf{Image Number} slider allows traversal of the images in the stack. The \textbf{Sync} button synchronizes images loaded in the stack to trajectory animations (activated by clicking the \textbf{Animate Trajectories} button in the \textbf{Visualization Panel} of the \textbf{Main Interface}.

\subsection{White Background}
\label{sec:white_background}

The user may change the default black background of the \textbf{Display Window} to white by toggling the \textbf{View > White Background} option.
\clearpage


\section{Performance}
\label{sec:performance}
This section outlines the performance and robustness of the inference technique subject to changes in calculation parameters. We emphasize that default calculation parameters are largely sufficient to yield accurate estimates of dynamic parameters from trajectories. For more extensive simulations and discussion of the performance of the technique, it is recommended that the user consult previous papers discussed in the \textbf{References} section. \\
\\
To give an overview of the performance, we use the diffusion as our test parameter. Other inferred parameters such as the forces, potential energy, and drift follow similar trends.

\subsection{Trajectory Length}
\label{sec:trajectory_length}

As discussed in Section \ref{sec:inference}, our trajectory inference technique does not depend on the length of trajectories. Equations \ref{eqn:d_inference}, \ref{eqn:df_inference}, \ref{eqn:ddrift_inference}, and \ref{eqn:dv_inference} clearly show that the posterior probability distribution is dependent on trajectory translocations (i.e., $\vec{r}^{k}_{\mu + 1} - \vec{r}^{k}_{\mu}$) and not trajectory lengths. Effectively, the ``long trajectory'' likelihood is simply the product of the individual likelihoods of each translocation.

\subsection{Number of Localizations}
\label{sec:number_of_localizations}

\begin{figure}[h!]
\center{\includegraphics[scale=0.1]{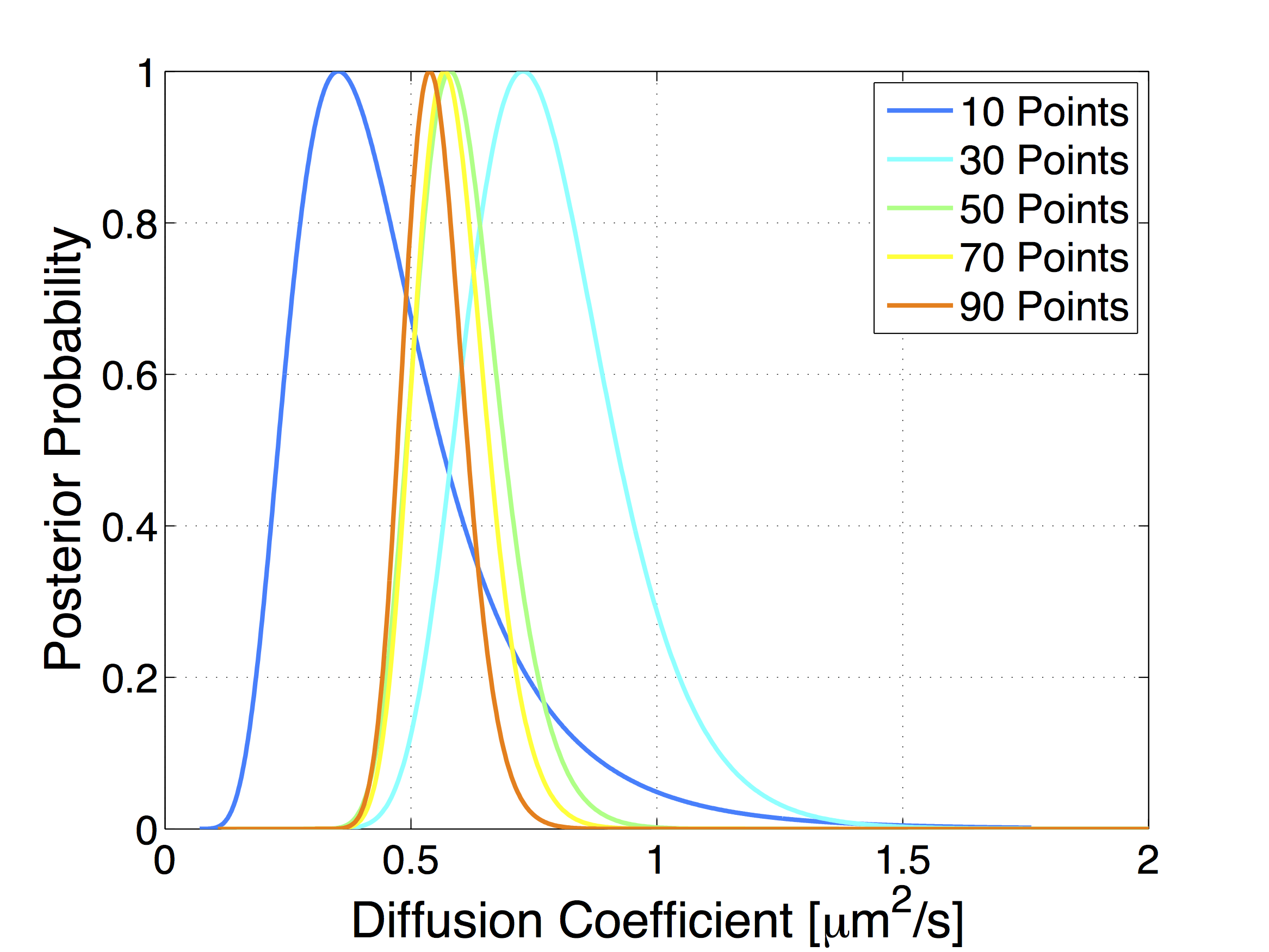}}
\caption{\label{fig:points_posteriors} Plots of posterior probabilities of the diffusion coefficient as a function of number of points (translocations) in the parameter estimation.}
\end{figure}

We show the effect of the number of localizations (translocations, effectively) on the inferred diffusion coefficient by sampling the posterior probability of the diffusion for increasingly long trajectories. This is entirely analogous to analyzing the posterior probability in a single zone of a mesh. This treatment is valid as per the logic of the previous section, as our technique uses individual translocations for parameter inference (i.e. it ``sees'' a long trajectory as a sum of two-point trajectories). \\
\\
Figure \ref{fig:points_posteriors} shows the form of the posterior for five different trajectories with different numbers of points (steps). Qualitatively, the distributions narrow with the increased number of points.

\begin{figure}[h!]
\center{\includegraphics[scale=0.1]{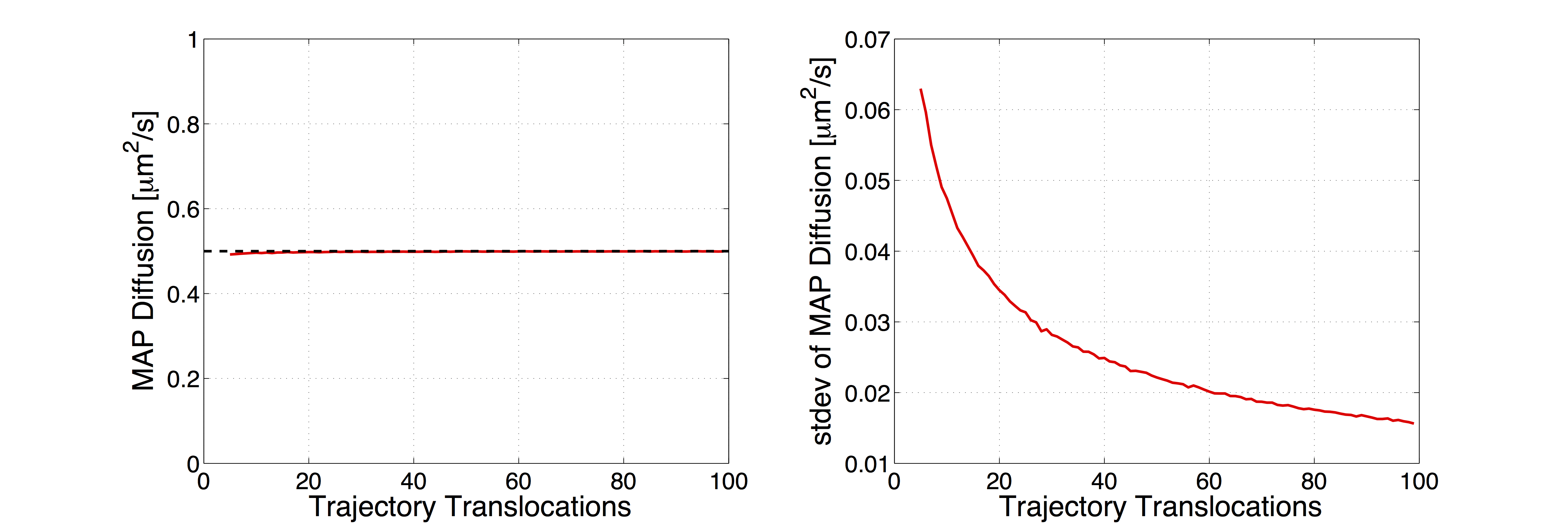}}
\caption{\label{fig:estimator} (left) Maximum a posteriori (MAP) estimator of diffusion coefficient on trajectories of increasing number of translocations (nominal diffusion of 0.5~$\mu m^2/s$). (right) Standard deviation of the MAP estimator.}
\end{figure}

The Bayesian inference estimator that we use is unbiased and we illustrate this in Figure \ref{fig:estimator}. Here, the nominal diffusion coefficient (0.5~$\mu m^2/s$) for trajectories of different numbers of translocations is estimated. Correspondingly, the error in the estimation is embodied by its standard deviation (stdev) which is described with a 1/$N$ profile. This is an important result, as the Bayesian inference method employed by \softwareName is unbiased regardless of the number of translocations per trajectory, or analogously, translocations per zone of a generated map.

\subsection{Mode Selection}
\label{sec:mode_selection}

\begin{figure}[h!]
\center{\includegraphics[scale=0.5]{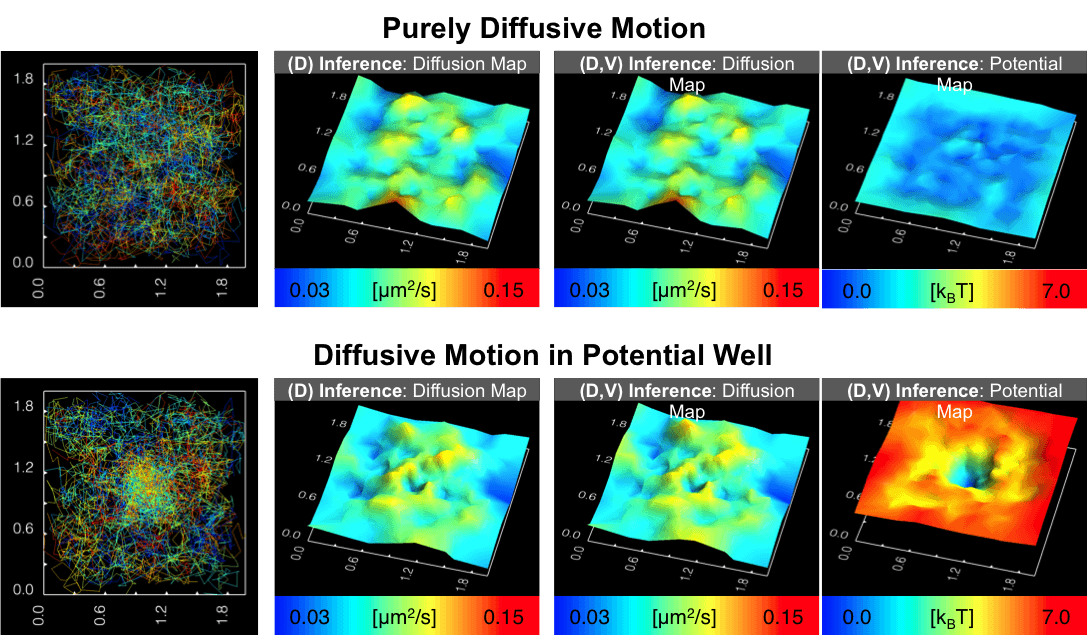}}
\caption{\label{fig:model_comparison} Comparison of different parameter maps inferred for the \textbf{(D) Inference} and \textbf{(D,V) Inference} modes.}
\end{figure}

\softwareName offers numerous modes to perform inference calculations to generate maps of dynamic parameters. This sections seeks to compare the \textbf{(D) Inference} and \textbf{(D,V) Inference} modes for two simulated datasets: one in which trajectory motion is purely diffusive, and another in which an interaction region is present. Effectively, two models of motion are being compared, and depending on which is selected key physical properties may be concealed. Results of the aforementioned comparison are displayed in Figure \ref{fig:model_comparison}.\\

The above examples clearly show that in the absence of an interaction energy (top row of Figure \ref{fig:model_comparison}), the corresponding potential energy map inferred using the \textbf{(D,V) Inference} mode is essentially flat, as is expected to be the case. Conversely, the \textbf{(D) Inference} mode does not model interaction energy, and is hence not capable of revealing a large interaction region (bottom row of Figure \ref{fig:model_comparison}).

\subsection{Prior Probabilities}
\label{sec:prior_probabilities}

\begin{figure}[h!]
\center{\includegraphics[scale=0.5]{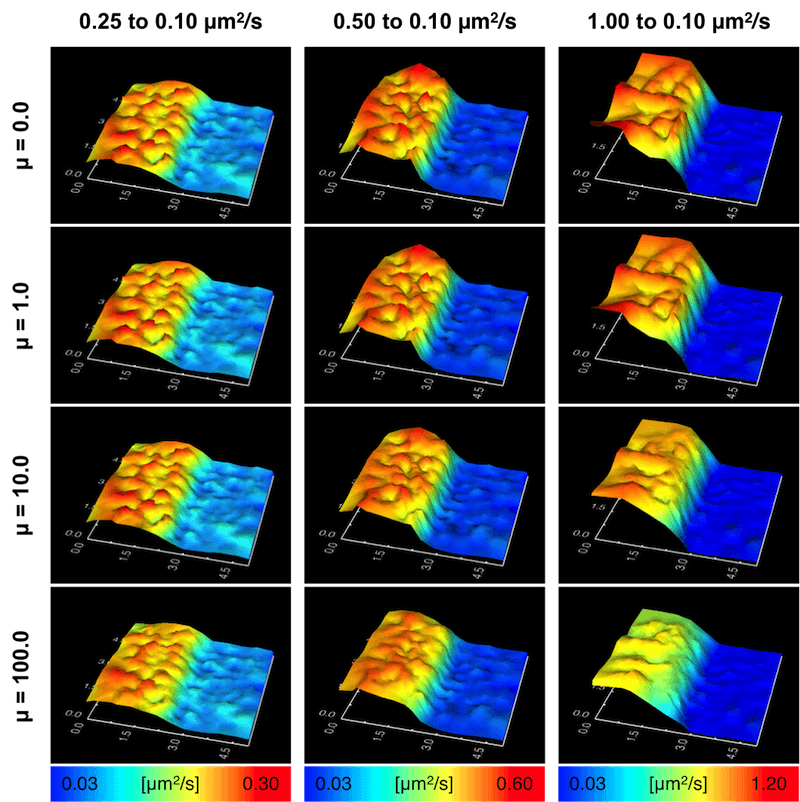}}
\caption{\label{fig:smoothing_prior_landscape} Table illustrating the effect of the diffusion map smoothing prior (based on the coefficient $\mu$) on simulated trajectories. The trajectory space has two diffusive populations separated by a ``step'' gradient of different sizes in the different columns.}
\end{figure}

\softwareName offers two types of prior probabilities. Jeffreys' prior is recommended in many cases. It is used to ensure that the posterior probability distribution of an inference is invariant by reparameterization. Moreover it protects the inference of diffusion in cases of high local confinement.\\
\\
The smoothing prior is used to penalize gradients in the inferred parameters between neighboring zones. To demonstrate the effect of the smoothing prior, we apply it to the case of a simulated trajectory set with different diffusion step gradients. Accordingly, the effect of the diffusion smoothing prior (i.e. the $\mu$ parameter described in Section \ref{sec:smoothing_prior}) on the diffusive map is illustrated in Figure \ref{fig:smoothing_prior_landscape}.

Increasing the value of $\mu$ smooths the interface between the two diffusive regions to different degrees. This prior is useful in cases in which large gradients in the diffusion are not to be expected. This is analogous to the use of the potential energy prior (i.e. the $\lambda$ parameter described in Section \ref{sec:smoothing_prior}).

\clearpage

\section{Stepped-Through Examples}
\label{sec:example}

In this section, a few examples for the different inference modes in \softwareName are stepped through in detail. All the examples are used with the included \emph{example} files, accessible from the \textbf{File} menu of the \textbf{Main Menu}.

\subsection{Example 1: Generating a Diffusion Map}
\label{sec:example_1}

This example demonstrates how to generate a diffusion map of simulated molecule trajectories with realistic diffusivities. The trajectory file is partitioned into four quadrants, each corresponding to a different specified diffusion coefficient. The localization precision (positioning noise) is specified to be 30~nm.

\begin{tabular}{ p{4.5cm} p{6cm} }
\multicolumn{2}{ p{\linewidth} }{\center\textbf{1.~Load Diffusion Example File}} \\
\small Select the \textbf{Diffusion Example (Simulation)} file from the \textbf{File > Examples} menu. The trajectories will start animating. The specified diffusion coefficient for the respective quadrants is overlaid as a background image.
& \raisebox{-\height}{\includegraphics[scale=0.35]{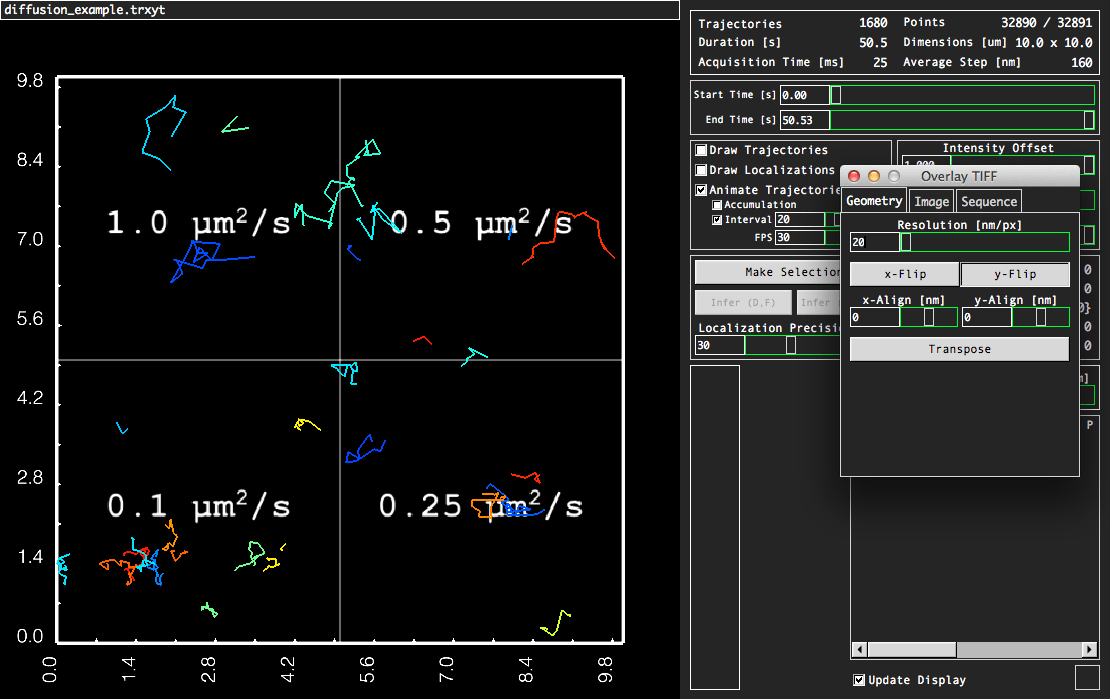}} \\
\end{tabular}

\begin{tabular}{ p{4.5cm} p{6cm} }
\multicolumn{2}{ p{\linewidth} }{\center\textbf{2.~Visualize Trajectories}} \\
\small Deselect the \textbf{Animate Trajectories} button, and select the \textbf{Draw Trajectories} button in the \textbf{Main Interface} to overlay all the trajectories in the file.
& \raisebox{-\height}{\includegraphics[scale=0.35]{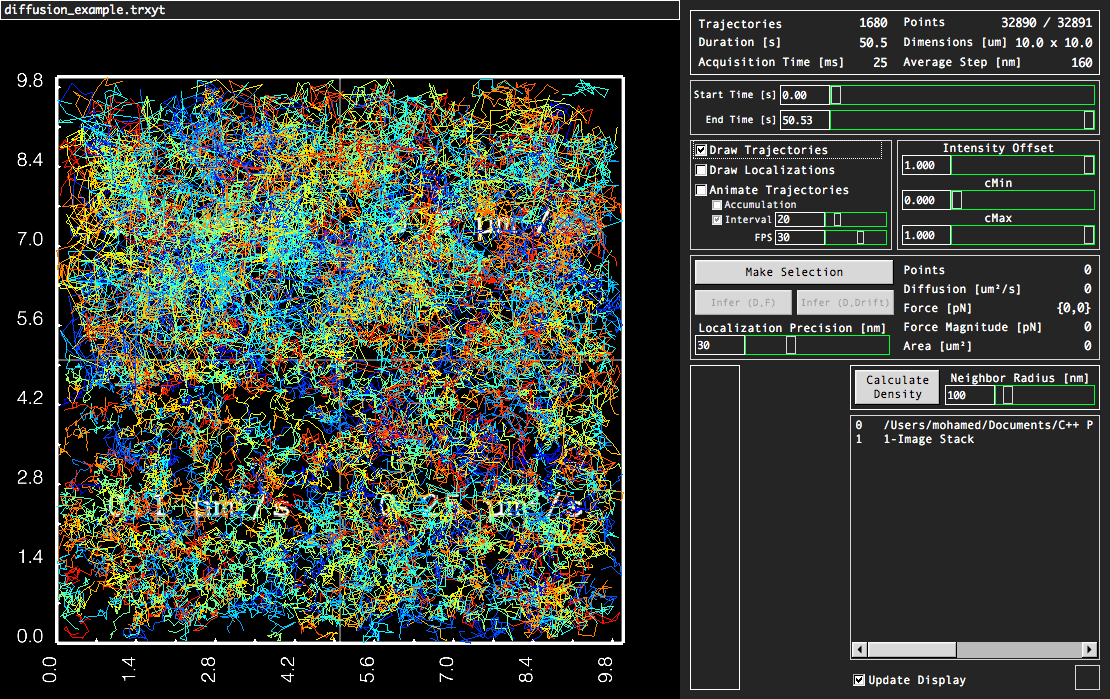}} \\
\end{tabular}

\begin{tabular}{ p{4.5cm} p{6cm} }
\multicolumn{2}{ p{\linewidth} }{\center\textbf{3.~Custom Selection Inference}} \\
\small The diffusion coefficient for a custom-selected zone can be estimated using the \textbf{Freehand Panel}. As the localization precision is already specified to 30~nm in the file, there is no need to adjust the \textbf{Localization Precision [nm]} slider. Select the \textbf{Make Selection} button, and draw a region inside the top left quadrant of the trajectory file by pressing and holding the left mouse button. Release the left mouse button to close the region. Now press the \textbf{Infer (D,F)} button to infer the diffusion and force inside the selected region. A diffusion value very close to the specified 1.0~$\mu m^{2}/s$ should be estimated (displayed in the right side of the \textbf{Freehand Panel}. Notice that the directional bias (indicated by the yellow arrow) is extremely weak in magnitude (indicated in the \textbf{Freehand Selection} panel, not by the size of the arrow!). This calculation can be redone for the different quadrants, giving accurate estimates of the specified diffusivity.
& \raisebox{-\height}{\includegraphics[scale=0.35]{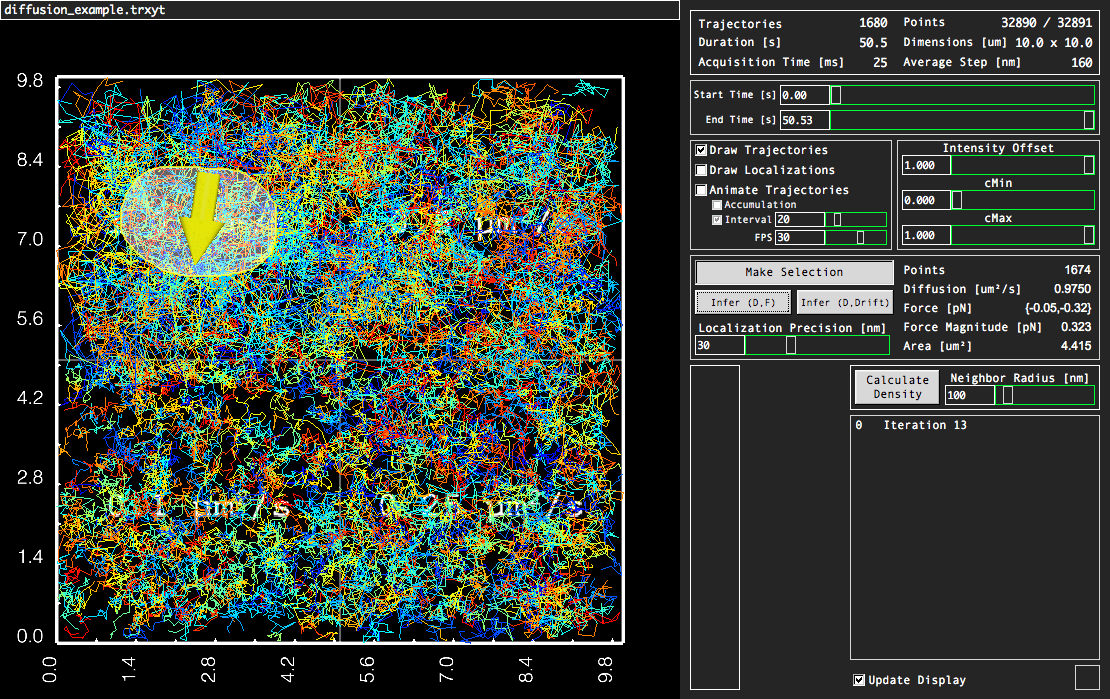}} \\
\end{tabular}

\begin{tabular}{ p{4.5cm} p{6cm} }
\multicolumn{2}{ p{\linewidth} }{\center\textbf{4.~Generate a Square Mesh}} \\
\small Deselect the \textbf{Make Selection} button in the \textbf{Main Interface}, and open the \textbf{Square Meshing} interface by accessing \textbf{File > Inference > Meshing > Square}. As we know the diffusivity to be relatively constant within each quadrant, we can choose relatively large grid spacings in the mesh. In the \textbf{Side Length [nm]} slider, manually type in 500~nm. Afterwards, press the \textbf{Apply} button to generate the mesh. The default colorcode corresponds to the number of points in each zone (seen in the \textbf{Colorbar}).
& \raisebox{-\height}{\includegraphics[scale=0.35]{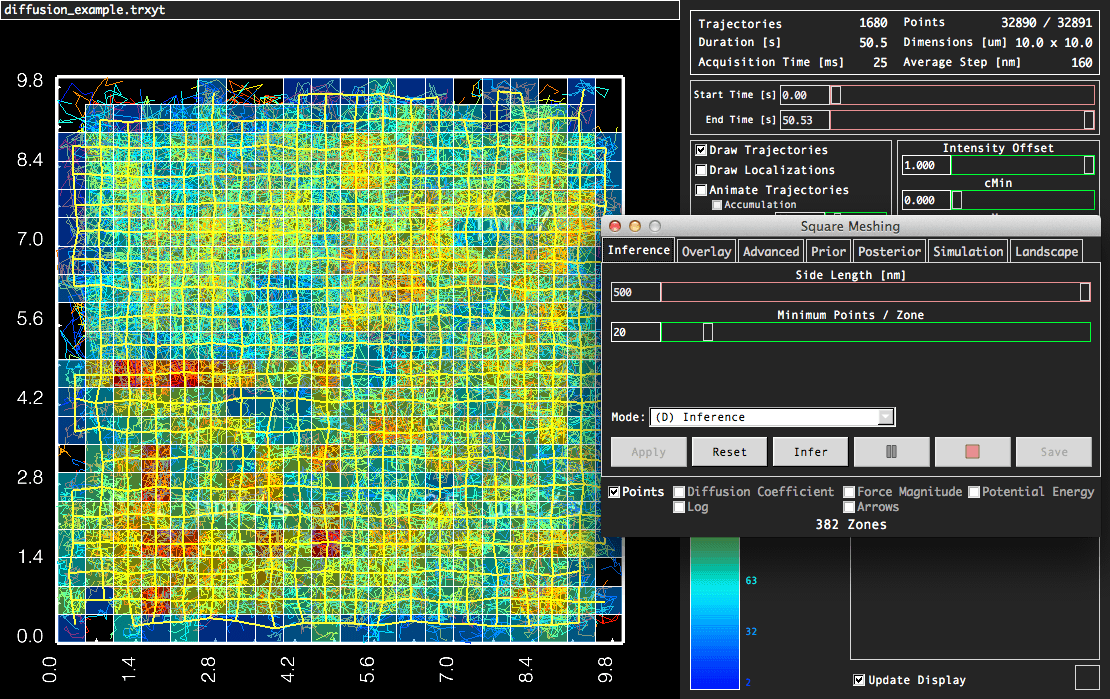}} \\
\end{tabular}

\begin{tabular}{ p{4.5cm} p{6cm} }
\multicolumn{2}{ p{\linewidth} }{\center\textbf{5.~Perform Inference}} \\
\small Make sure that the \textbf{(D) Inference} is selected in the \textbf{Mode} drop down menu in the meshing interface. The localization precision may be adjusted in the \textbf{Advanced Tab} of the meshing interface, however, its default value is set to 30~nm which corresponds to the value in the trajectory file. Press the \textbf{Infer} button to infer the diffusion coefficient in each of the zones in the mesh. The \textbf{Colorbar} will automatically update to the diffusion values. Notice that values in each of the zones accurately corresponds to the diffusion specified in the file.
& \raisebox{-\height}{\includegraphics[scale=0.35]{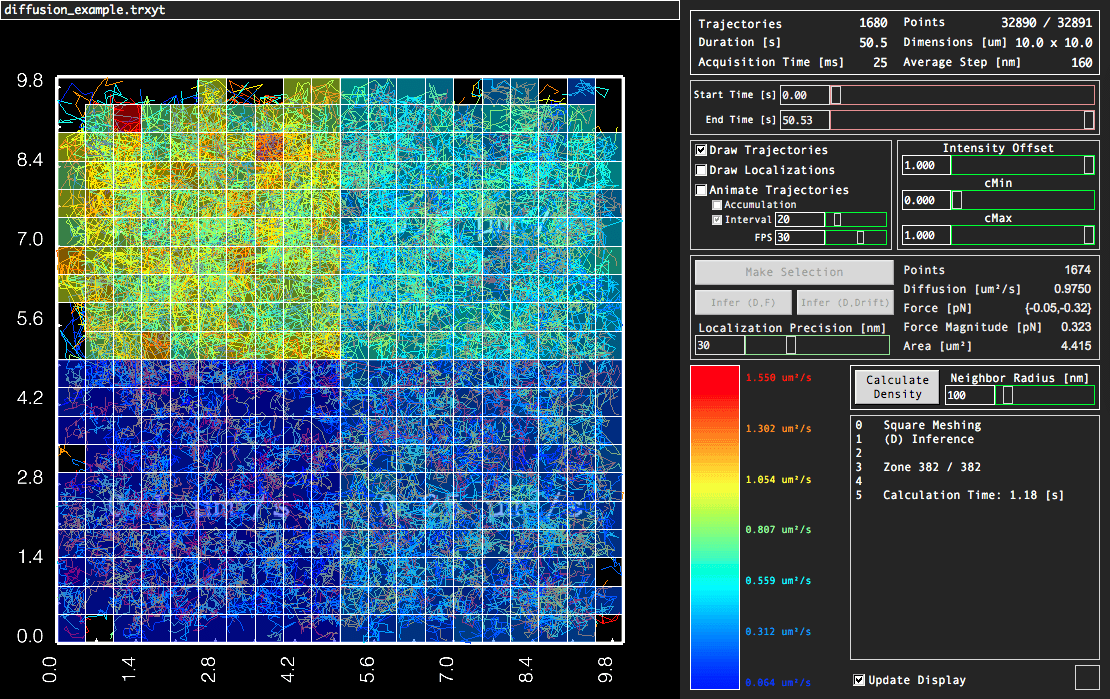}} \\
\end{tabular}

\begin{tabular}{ p{4.5cm} p{6cm} }
\multicolumn{2}{ p{\linewidth} }{\center\textbf{6.~Landscape Viewing}} \\
\small For more intuitive visualization, the diffusion map can be viewed in a 3D landscape mode. Select the \textbf{Landscape} tab in the meshing interface, and press \textbf{View Landscape}. The \textbf{Display Window} becomes manipulatable in 3D, clearly showing the differences in diffusion between the different quadrants. Sliders in the \textbf{Landscape} tab can be used to adjust some of the landscape display properties.
& \raisebox{-\height}{\includegraphics[scale=0.35]{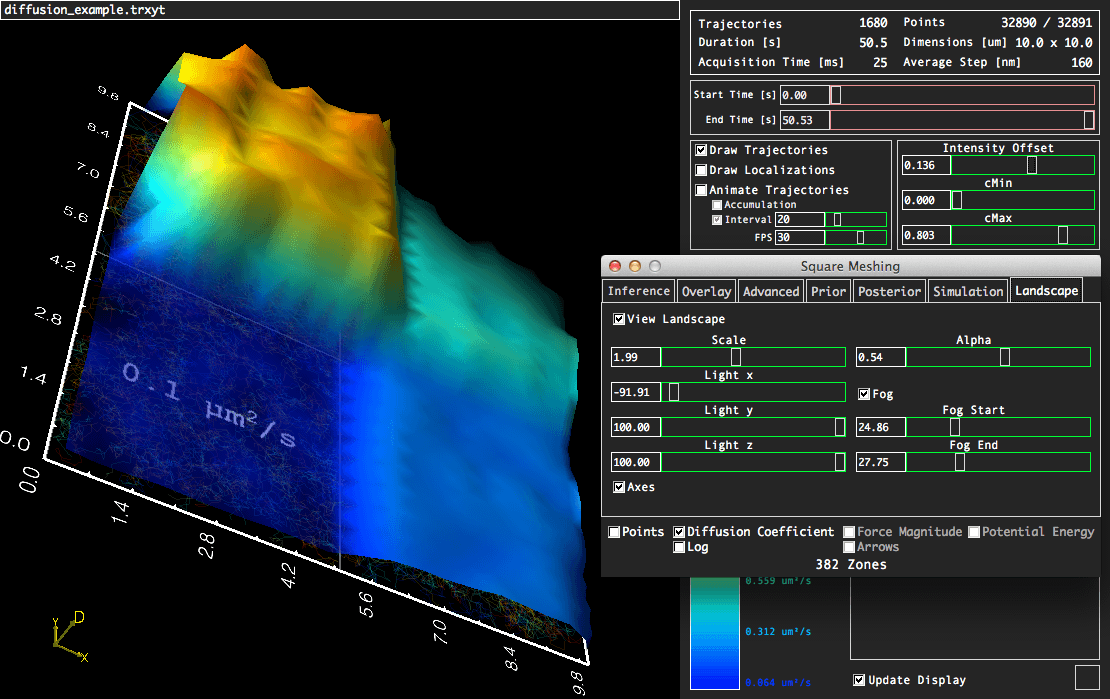}} \\
\end{tabular}

\clearpage

\subsection{Example 2: Generating a Potential (Interaction) Energy Map}
\label{sec:example_2}

This example demonstrates how to generate a potential (interaction) energy map of simulated molecule trajectories with realistic diffusivities. The trajectory file is partitioned into four quadrants, each containing a potential energy well with a different depth (the specified depth value is indicated in the background image). The form of each well is Gaussian, with a variance of 300~nm. The localization precision for the trajectories is specified to be 30~nm. The trajectories are set to diffuse at $0.2 \mu m^{2}/s$.

\begin{tabular}{ p{4.5cm} p{6cm} }
\multicolumn{2}{ p{\linewidth} }{\center\textbf{1.~Load Potential Example File}} \\
\small Select the \textbf{Potential Example (Simulation)} file from the \textbf{File > Examples} menu. The trajectories will start animating. The specified potential energy depths for the respective quadrants is overlaid as a background image.
& \raisebox{-\height}{\includegraphics[scale=0.35]{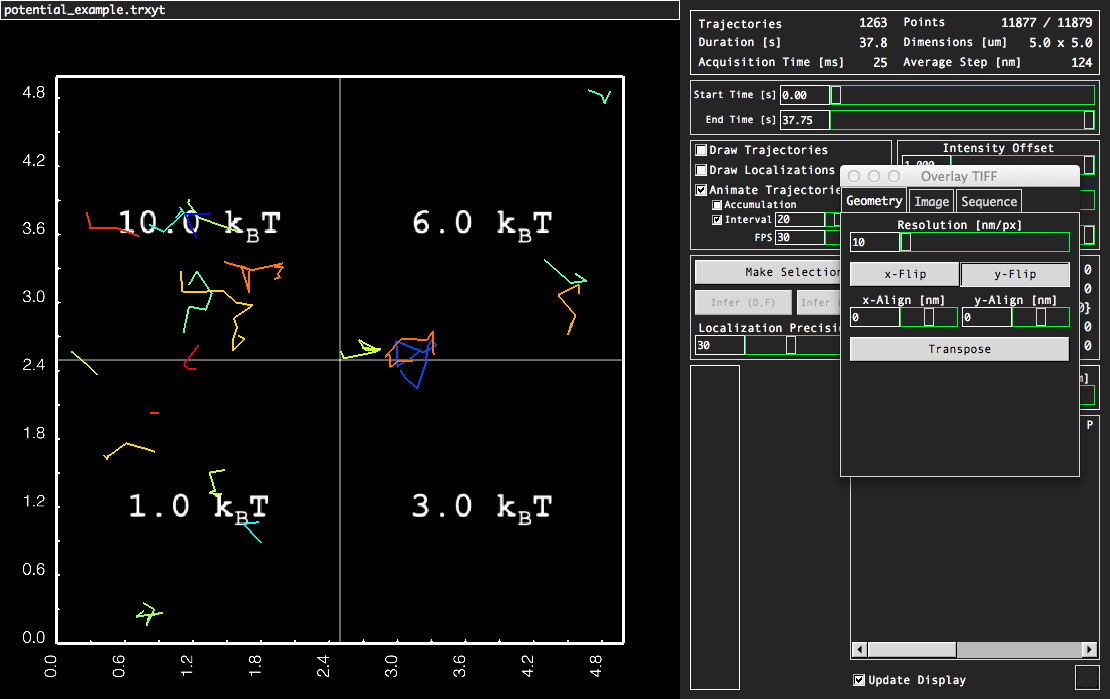}} \\
\end{tabular}

\begin{tabular}{ p{4.5cm} p{6cm} }
\multicolumn{2}{ p{\linewidth} }{\center\textbf{2.~Visualize Trajectories}} \\
\small Deselect the \textbf{Animate Trajectories} button, and select the \textbf{Draw Trajectories} button in the \textbf{Main Interface} to overlay all the trajectories in the file.
& \raisebox{-\height}{\includegraphics[scale=0.35]{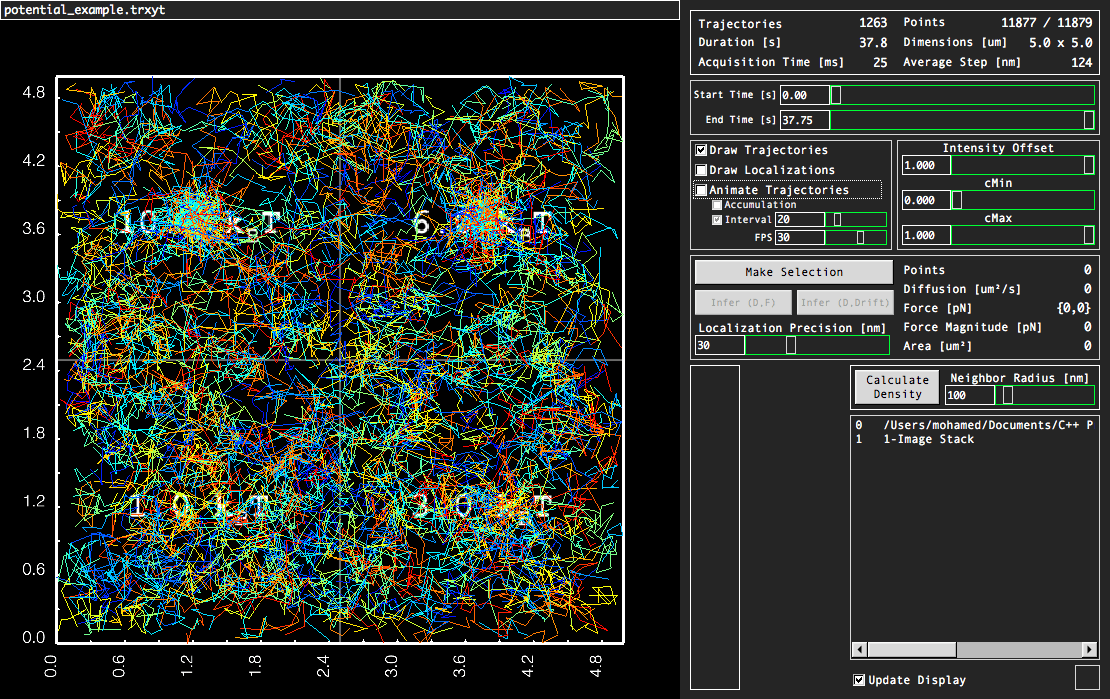}} \\
\end{tabular}

\begin{tabular}{ p{4.5cm} p{6cm} }
\multicolumn{2}{ p{\linewidth} }{\center\textbf{3.~Calculate Density}} \\
\small To help reveal regions of attraction, we can perform a naive density calculation to show where trajectory points are most concentrated. Press the \textbf{Calculate Density} button in the \textbf{Density Panel} of the \textbf{Main Interface} (here we leave the \textbf{Neighborhood Radius} to its default value. Upon calculation, the colorcode will correspond to the relative density of each localization in the loaded trajectory file, red being the highest and dark blue being the lowest.
& \raisebox{-\height}{\includegraphics[scale=0.35]{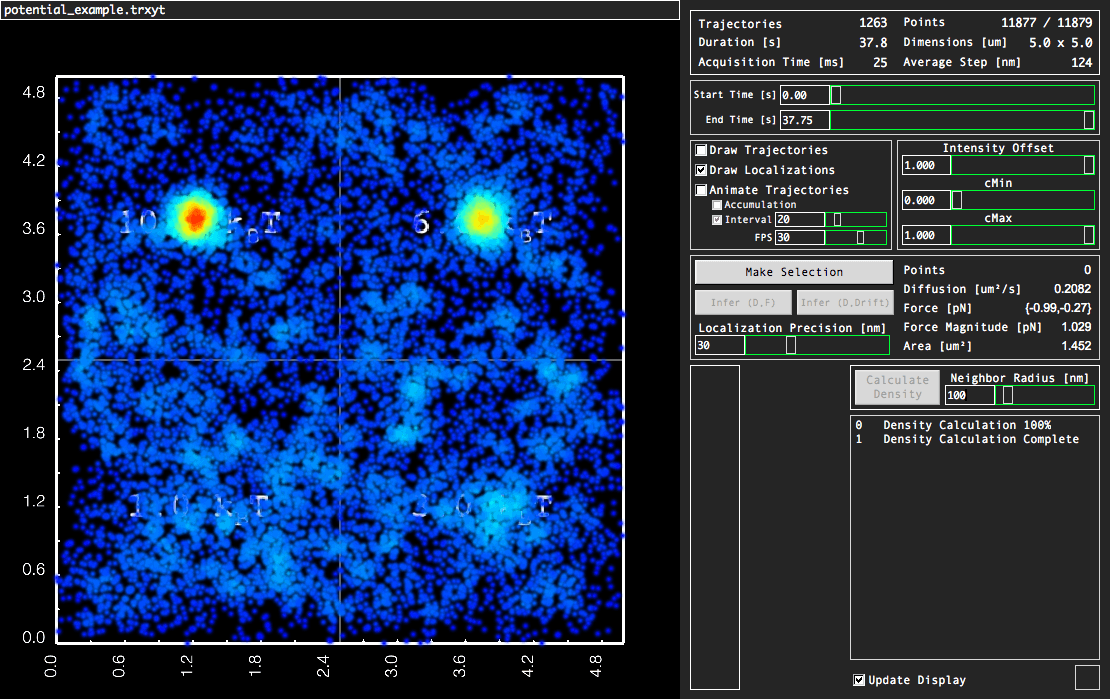}} \\
\end{tabular}

\begin{tabular}{ p{4.5cm} p{6cm} }
\multicolumn{2}{ p{\linewidth} }{\center\textbf{3.~Custom Selection Inference}} \\
\small Before generate the potential energy map, we can infer directional bias of trajectories at the proximity of high density localization regions. The force for a custom-selected zone can be estimated using the \textbf{Freehand Panel}. As the localization precision is already specified to 30~nm in the file, there is no need to adjust the \textbf{Localization Precision [nm]} slider. Select the \textbf{Make Selection} button, and draw a region adjacent to a high density region revealed by the density calculation. Now press the \textbf{Infer (D,F)} button to infer the diffusion and force inside the selected region. For the deepest wells, the force is strong in magnitude and strongly directional towards the well (displayed in the right side of the \textbf{Freehand Panel}. The more shallow wells will show weaker tendencies.
& \raisebox{-\height}{\includegraphics[scale=0.35]{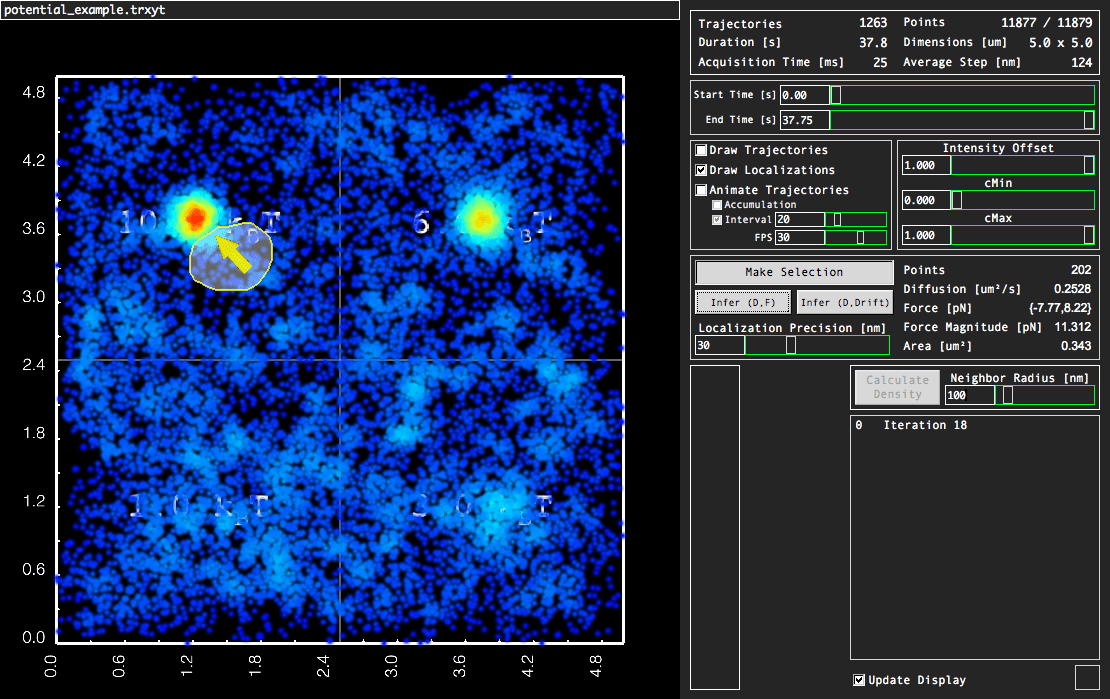}} \\
\end{tabular}

\begin{tabular}{ p{4.5cm} p{6cm} }
\multicolumn{2}{ p{\linewidth} }{\center\textbf{5.~Generate a Quad-Tree Mesh}} \\
\small Deselect the \textbf{Make Selection} button in the \textbf{Main Interface}, and open the \textbf{Quad-Tree Meshing} interface by accessing \textbf{File > Inference > Meshing > Quad-Tree}.  Here we opt for the default meshing values, although the user is encouraged to test different values. Press the \textbf{Apply} button to generate the mesh. The yellow lines connecting each of the zones corresponds to which neighboring zones ``see each other'' (Described in Section \ref{sec:neighboring_zone_connections}). The default colorcode corresponds to the number of points in each zone (seen in the \textbf{Colorbar}).
& \raisebox{-\height}{\includegraphics[scale=0.35]{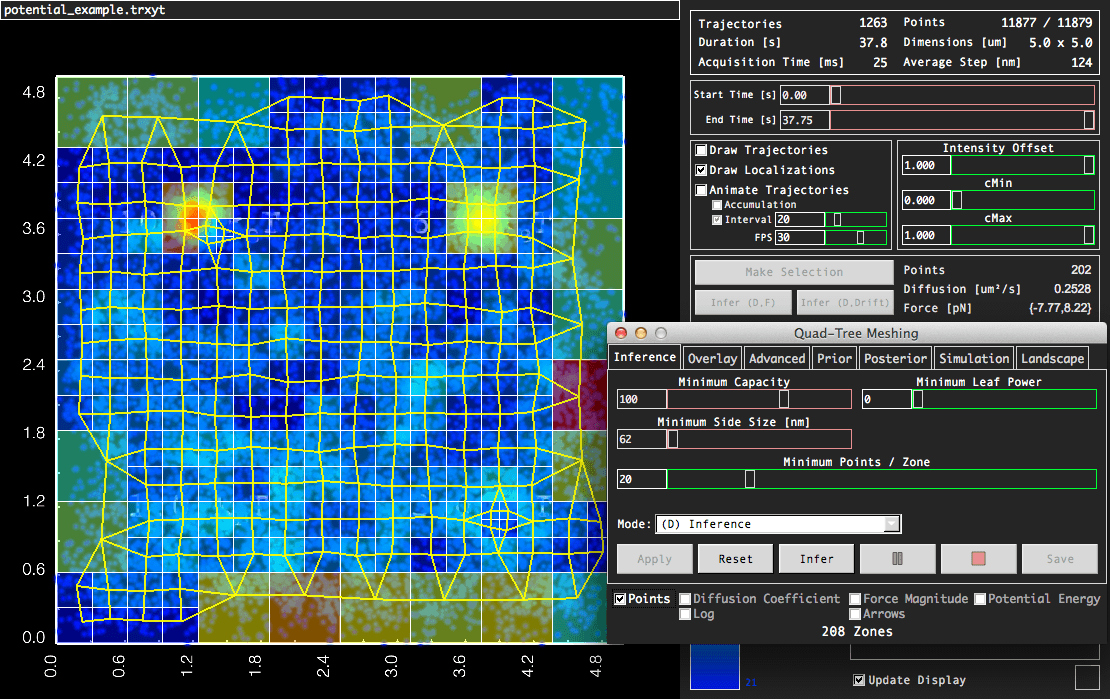}} \\
\end{tabular}

\begin{tabular}{ p{4.5cm} p{6cm} }
\multicolumn{2}{ p{\linewidth} }{\center\textbf{6.~Setup a Randomized Optimization Calculation}} \\
\small Select the \textbf{(D,V) Inference} option from the \textbf{Mode} drop down menu in the meshing interface. The localization precision may be adjusted in the \textbf{Advanced Tab} of the meshing interface, however, its default value is set to 30~nm which corresponds to the value in the trajectory file. As the \textbf{(D,V) Inference} calculation is quite computationally expensive, we choose to perform a randomized optimization which greatly reduces the time needed to infer parameters from the trajectories. In the \textbf{Advanced} tab, select the \textbf{Randomized Optimization} button, and increase the \textbf{Selection Radius [nm]} slider to roughly 1000~nm and reduce the \textbf{Cost Tolerance [\%]} slider to zero (implying we will have to manually stop the calculation).
& \raisebox{-\height}{\includegraphics[scale=0.35]{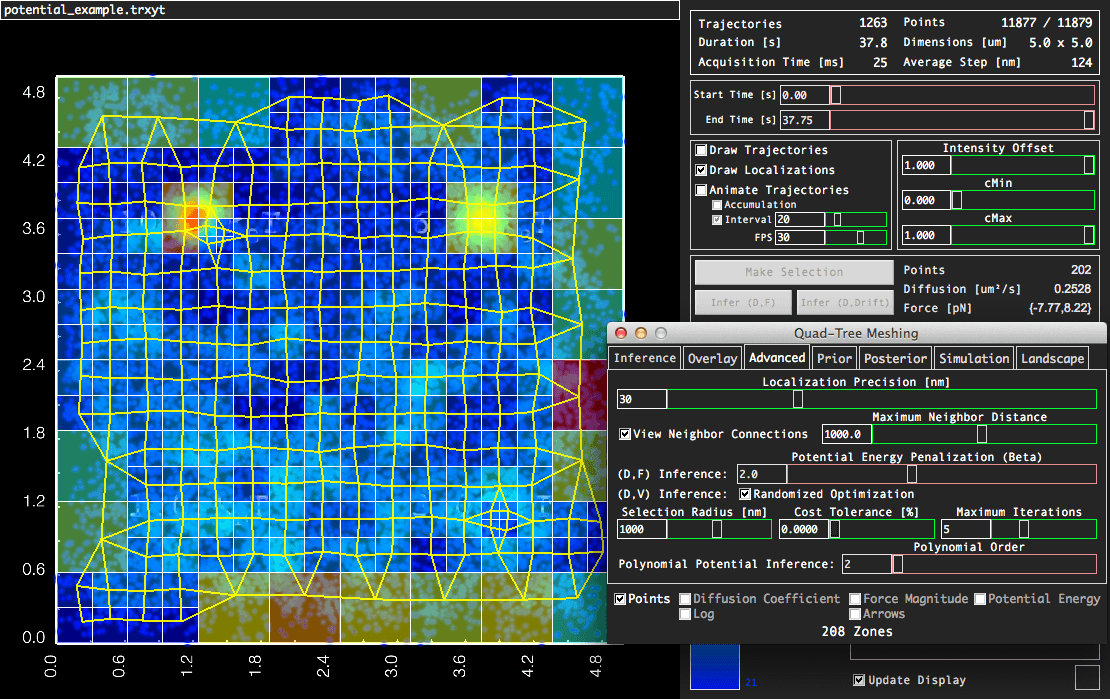}} \\
\end{tabular}

\begin{tabular}{ p{4.5cm} p{6cm} }
\multicolumn{2}{ p{\linewidth} }{\center\textbf{7.~Perform Inference}} \\
\small Select the \textbf{Inference} tab and press the \textbf{Infer} button to infer the diffusion, forces, and potential energy in each of the zones in the mesh using randomized optimization. Subregions of the mesh appearing white will appear white indicating that they are being optimized. The idea is to perform the randomized optimization until the cost function has sufficiently decayed. In this example, this corresponds to approximately 200 iterations.
& \raisebox{-\height}{\includegraphics[scale=0.35]{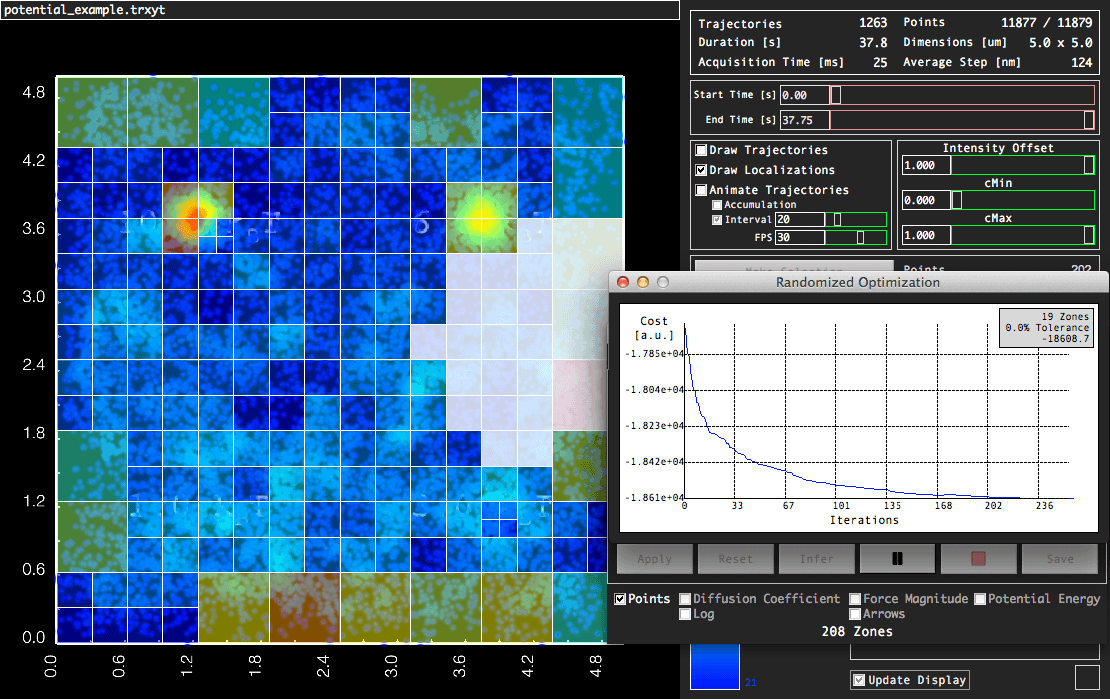}} \\
\end{tabular}

\begin{tabular}{ p{4.5cm} p{6cm} }
\multicolumn{2}{ p{\linewidth} }{\center\textbf{8.~Stop Calculation}} \\
At roughly 200 iterations in the randomized optimization (indicated in the \textbf{Randomized Optimization} window), press the \textbf{Pause} (double verticle lines) button then the \textbf{Stop} (red square) to stop the inference calculation. Press the \textbf{Potential Energy} button to overlay the potential energy values to the mesh, with yellow arrows corresponding to the force direction overlaid. Clearly seen from the yellow force arrows is the directional bias of the trajectories. In fact, for the deep wells will have a strong bias, whereas the weaker ones will not. In the bottom left quadrant, the depth of the potential well corresponds to $1.0~k_{B}T$ which is roughly the background potential energy due to thermal fluctuations.
& \raisebox{-\height}{\includegraphics[scale=0.35]{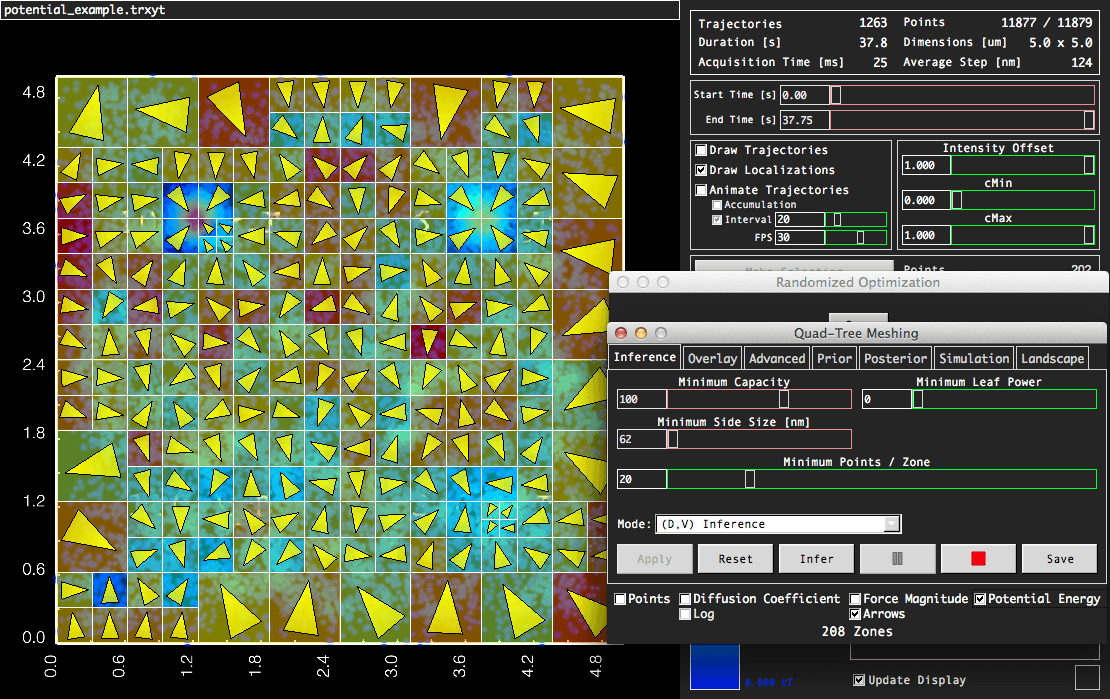}} \\
\end{tabular}

\begin{tabular}{ p{4.5cm} p{6cm} }
\multicolumn{2}{ p{\linewidth} }{\center\textbf{9.~Landscape Viewing}} \\
\small For more intuitive visualization, the potential energy map can be viewed in a 3D landscape mode. Select the \textbf{Landscape} tab in the meshing interface, and press \textbf{View Landscape}. The \textbf{Display Window} becomes manipulatable in 3D, clearly showing the different potential energy wells between the different quadrants. Sliders in the \textbf{Landscape} tab can be used to adjust some of the landscape display properties. Here, we see that the approximate depths of the potential energy wells correspond to the specified values.
& \raisebox{-\height}{\includegraphics[scale=0.35]{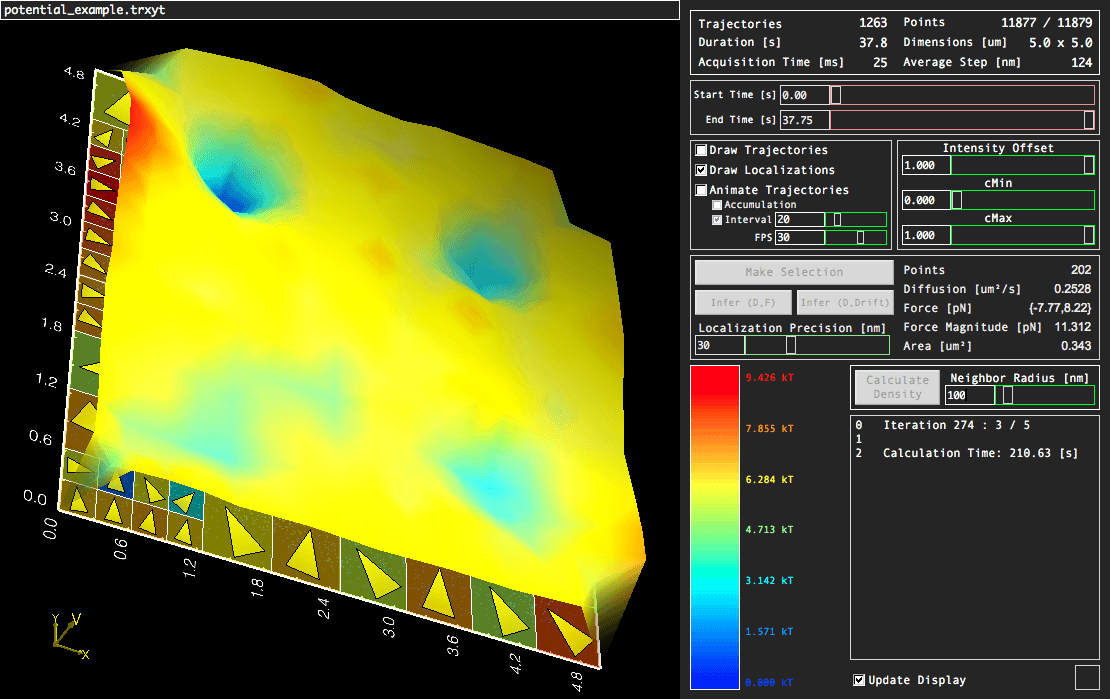}} \\
\end{tabular}

\clearpage

\subsection{Example 3: Membrane Microdomain}
\label{sec:example_3}

This example demonstrates how to generate a potential (interaction) energy map based on the trajectory of an $\epsilon$-toxin receptor tagged with an amine-coated lanthanide oxide nanoparticle on an MDCK cell \cite{turkcan2012_2}. The receptor is seen to hop between three different lipid rafts, each of which having a different strength of confinement. Experimentally, these trajectories were captured in a wide field fluorescent microscope configuration.

\begin{tabular}{ p{4.5cm} p{6cm} }
\multicolumn{2}{ p{\linewidth} }{\center\textbf{1.~Load Toxin Receptor Example File}} \\
\small Select the \textbf{Toxin Receptor in Lipid Raft)} file from the \textbf{File > Examples} menu.
& \raisebox{-\height}{\includegraphics[scale=0.35]{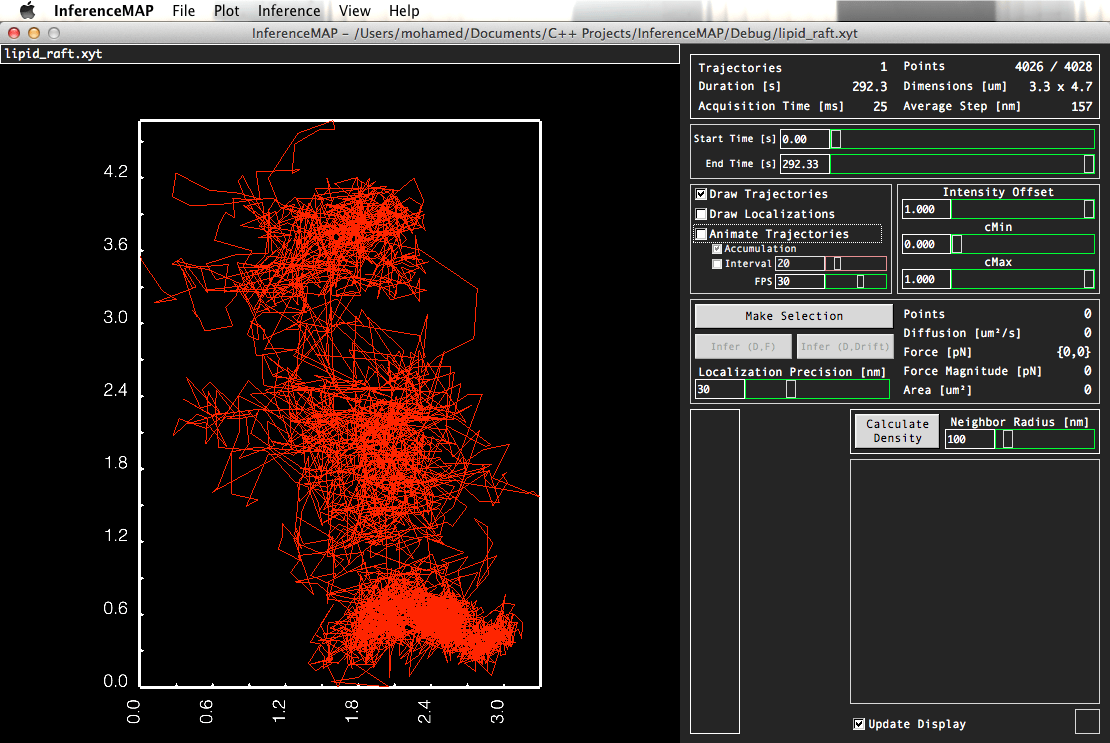}} \\
\end{tabular}

\begin{tabular}{ p{4.5cm} p{6cm} }
\multicolumn{2}{ p{\linewidth} }{\center\textbf{2.~Generate a Voronoi Mesh}} \\
\small Open the \textbf{Voronoi Meshing} interface from the \textbf{File > Inference > Meshing > Voronoi} menu. For this meshing mode, the number of zones must be predefined beforehand. \softwareName automatically selects an appropriate number based on the number of localizations in the loaded data set. Here we shall use the default value of 50 zones. Press the \textbf{Apply} button to generate the Voronoi mesh.
& \raisebox{-\height}{\includegraphics[scale=0.35]{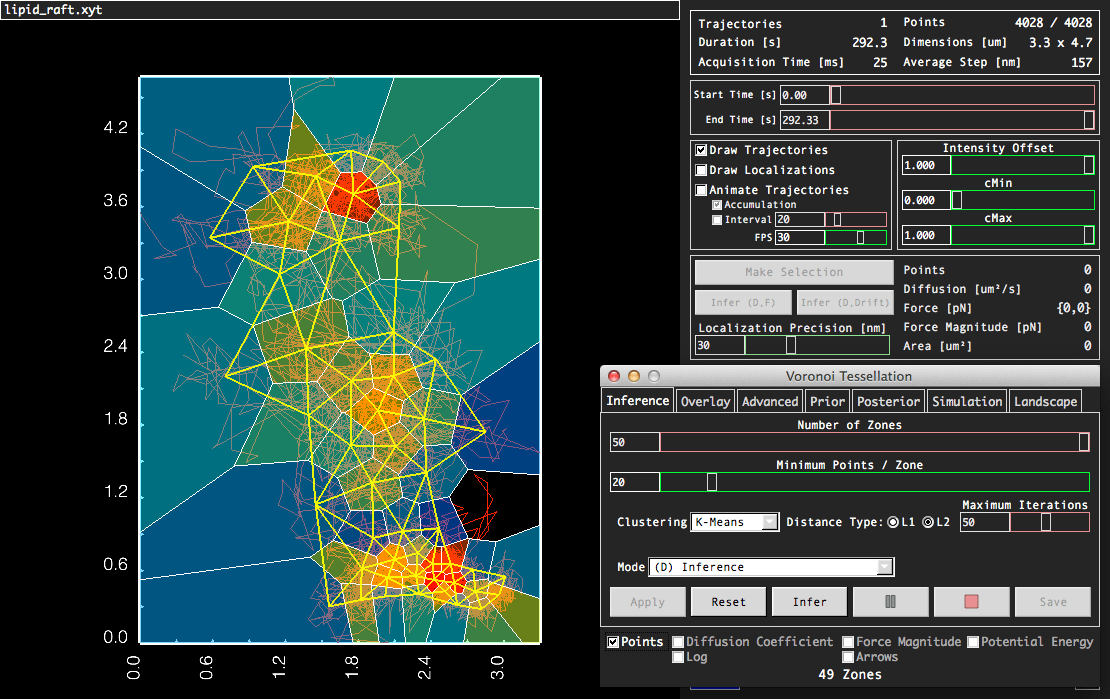}} \\
\end{tabular}

\begin{tabular}{ p{4.5cm} p{6cm} }
\multicolumn{2}{ p{\linewidth} }{\center\textbf{3.~Adjust Mesh}} \\
\small The generated mesh is seen to possess a deactivated zone in the bottom right corner. Activate this zone by right clicking on it. The yellow lines connecting each of the zones corresponds to which neighboring zones ``see each other'' (Described in Section \ref{sec:neighboring_zone_connections}). In this case, some of the connections between zones do not make sense based on the layout of the trajectory (e.g. the top two rafts do not seem to be associated). Select the \textbf{Advanced} tab and adjust \textbf{Maximum Neighbor Distance [nm]} to roughly 715~nm. Additionally, adjust the \textbf{Localization Precision [nm]} slider to 20~nm (corresponding to the approximate experimental value).
& \raisebox{-\height}{\includegraphics[scale=0.35]{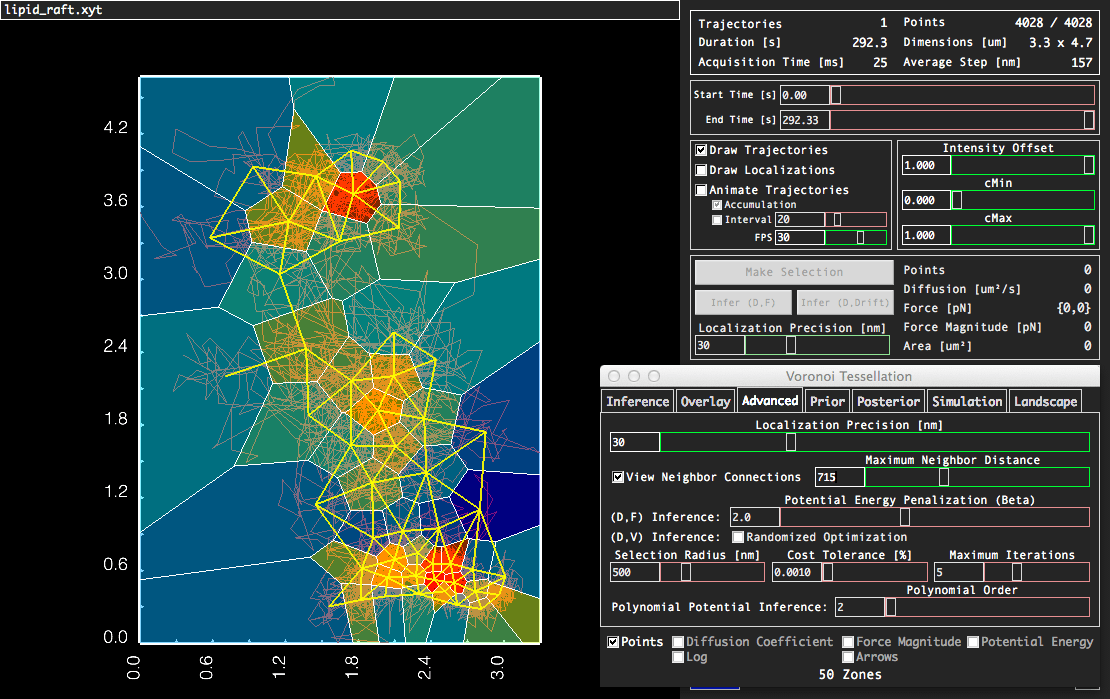}} \\
\end{tabular}

\begin{tabular}{ p{4.5cm} p{6cm} }
\multicolumn{2}{ p{\linewidth} }{\center\textbf{4.~Perform Inference}} \\
\small In the \textbf{Inference} tab, select the \textbf{(D,V) Inference} mode. To perform the inference calculation, press \textbf{Infer}. The \textbf{(D,V) Inference} mode is generally a slow and expensive calculation and necessitates the randomized optimization feature to complete in a reasonable amount of time. However, as there are only 50 zones in this mesh, the calculation time will not be too lengthy (roughly 2 minutes).
& \raisebox{-\height}{\includegraphics[scale=0.35]{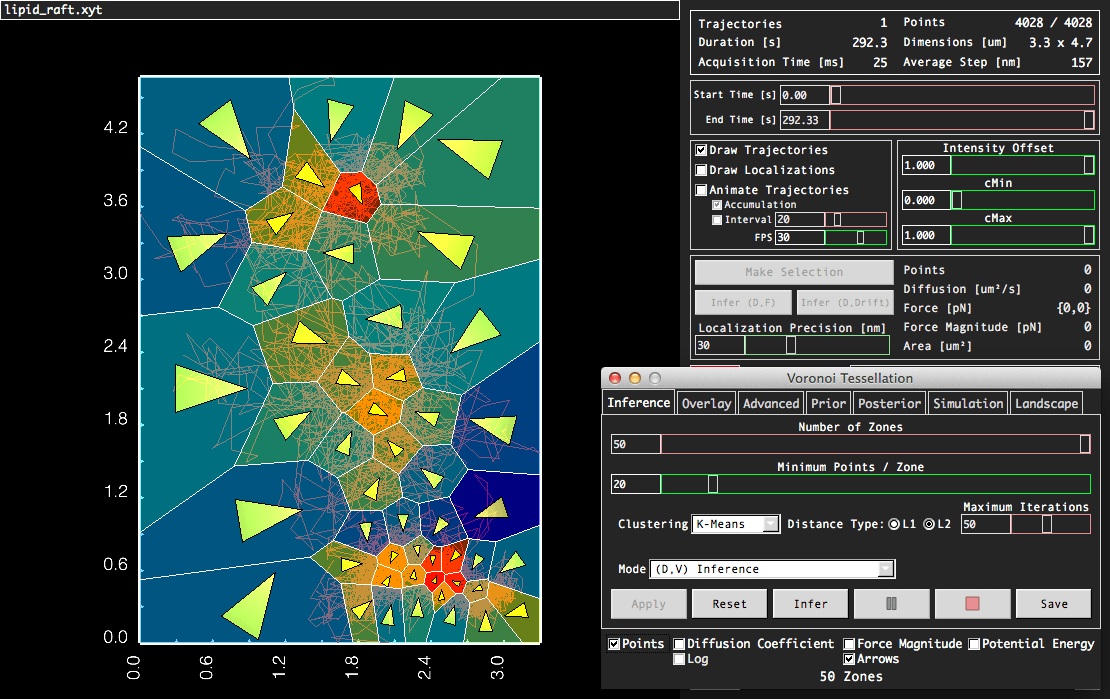}} \\
\end{tabular}

\begin{tabular}{ p{4.5cm} p{6cm} }
\multicolumn{2}{ p{\linewidth} }{\center\textbf{5.~Landscape Viewing}} \\
\small Based on the force arrows it is seen that the toxin receptor is separately confined in three distinct parts of the trajectory. To see the potential energy field which gives rise to this confining force, select \textbf{Potential Energy} in the bottom of the meshing interface. Adjust the \textbf{cMin} and \textbf{cMax} sliders in the \textbf{Main Interface} to 0.2 and 0.8, respectively. Now select the \textbf{Overlay} tab, and choose \textbf{Blue Red} \textbf{Map} drop-down menu. Next, select the \textbf{Landscape} tab and press \textbf{View Landscape}. The \textbf{Display Window} becomes manipulatable in 3D. Represented are three potential energy wells, the bottom most one being much deeper (roughly 5~$k_{B}T$) than the top two (roughly 2~$k_{B}T$). Essentially, the bottom strongly stabilizes the toxin receptor, while the top two wells are too shallow to keep the receptor confined for long.
& \raisebox{-\height}{\includegraphics[scale=0.35]{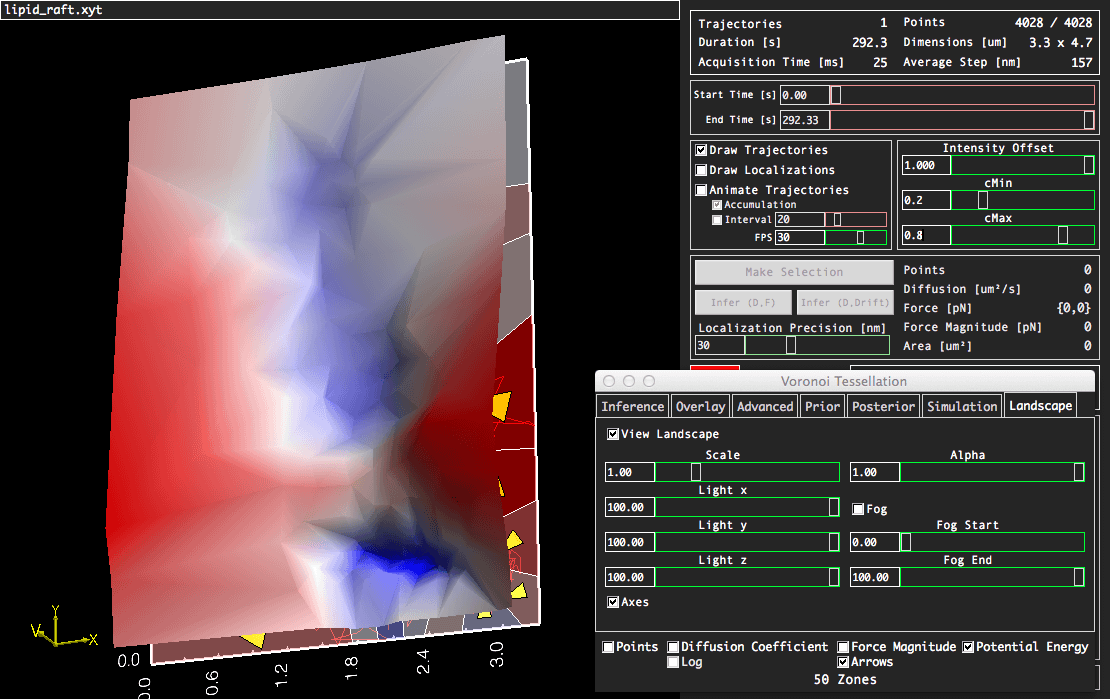}} \\
\end{tabular}

\clearpage

\subsection{Example 4: Neurotransmitter Receptors}
\label{sec:example_4}

This example demonstrates how to create a diffusion map for glycine receptors on a mouse hippocampal neuron membrane. Experimentally, these trajectories were captured in a uPAINT configuration, in a similar fashion to \cite{masson2014}.

\begin{tabular}{ p{4.5cm} p{6cm} }
\multicolumn{2}{ p{\linewidth} }{\center\textbf{1.~Load Neurotransmitter Receptor File}} \\
\small Select the \textbf{Glycine Receptors on Mouse Hippocampal Neuron)} file from the \textbf{File > Examples} menu. The trajectories will start animating in an accumulation mode, demonstrating how rapidly the surface of the apical membrane is explored by the tracked glycine receptors. A GFP image is overlaid to show the shape of the neuron.
& \raisebox{-\height}{\includegraphics[scale=0.35]{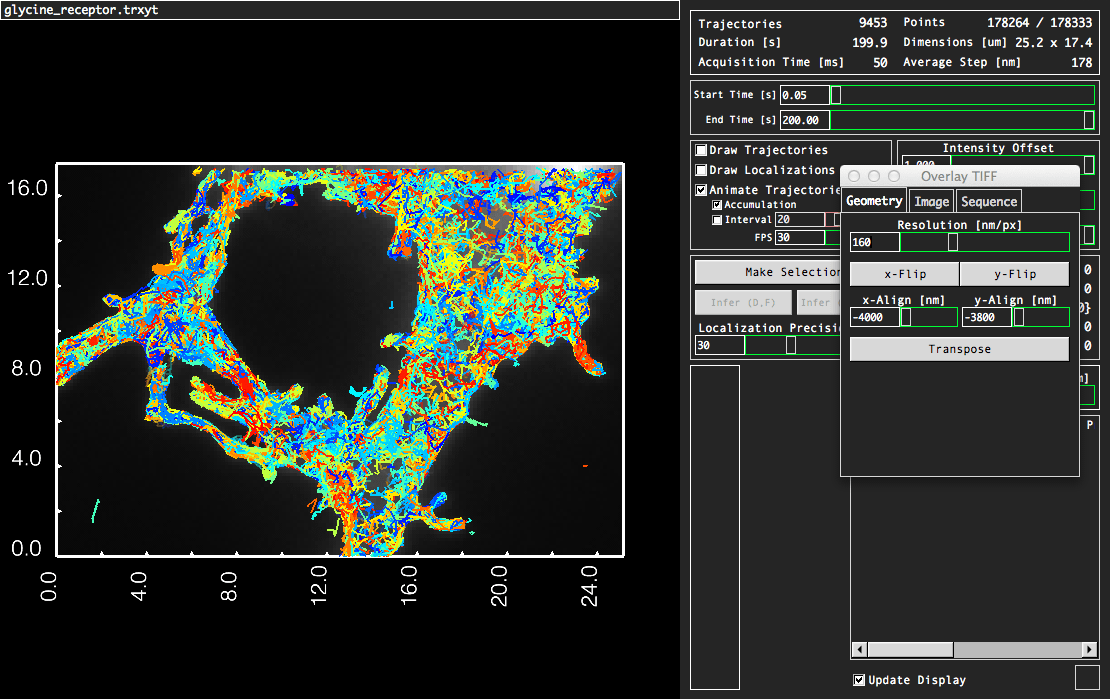}} \\
\end{tabular}

\begin{tabular}{ p{4.5cm} p{6cm} }
\multicolumn{2}{ p{\linewidth} }{\center\textbf{2.~Visualize Trajectories}} \\
\small Deselect the \textbf{Animate Trajectories} button, and select the \textbf{Draw Trajectories} button in the \textbf{Main Interface} to overlay all the trajectories in the file.
& \raisebox{-\height}{\includegraphics[scale=0.35]{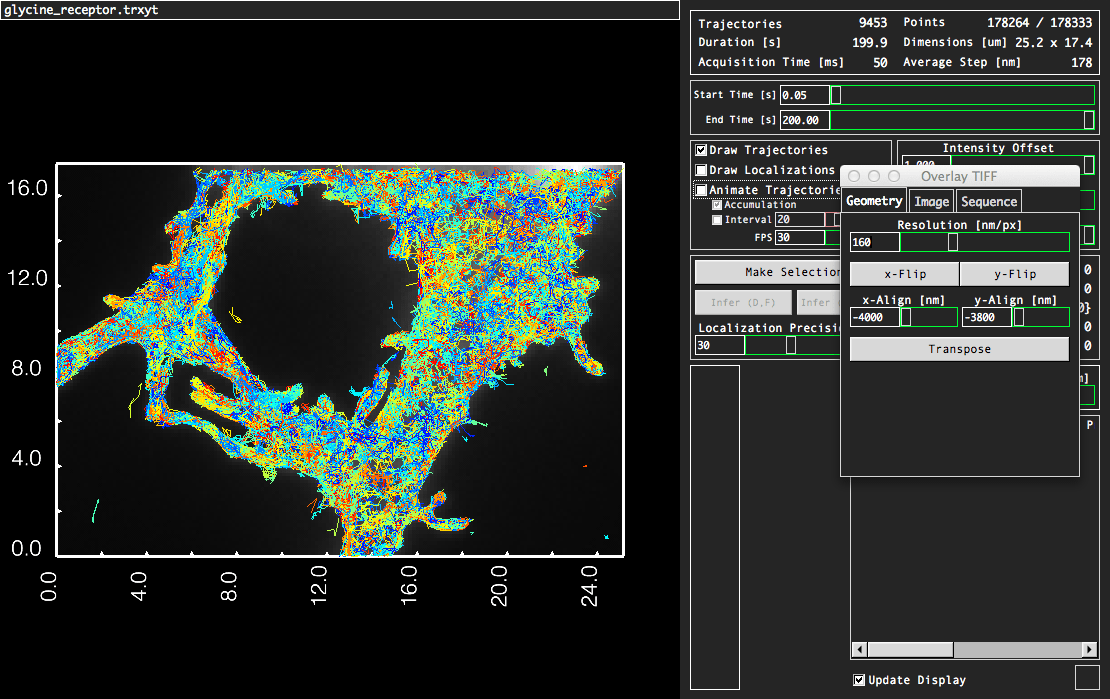}} \\
\end{tabular}

\begin{tabular}{ p{4.5cm} p{6cm} }
\multicolumn{2}{ p{\linewidth} }{\center\textbf{3.~Make Custom Selection}} \\
\small Zoom into the region to the left by right clicking and dragging the mouse to generate a box in the \textbf{Display Window}. Letting go will zoom into the selected region (double-clicking will reset the view if you wish to reselect). Press the \textbf{Make Selection} button in the \textbf{Freehand Panel}, and select a region surrounding the part of the neuron indicated in the figure.
& \raisebox{-\height}{\includegraphics[scale=0.35]{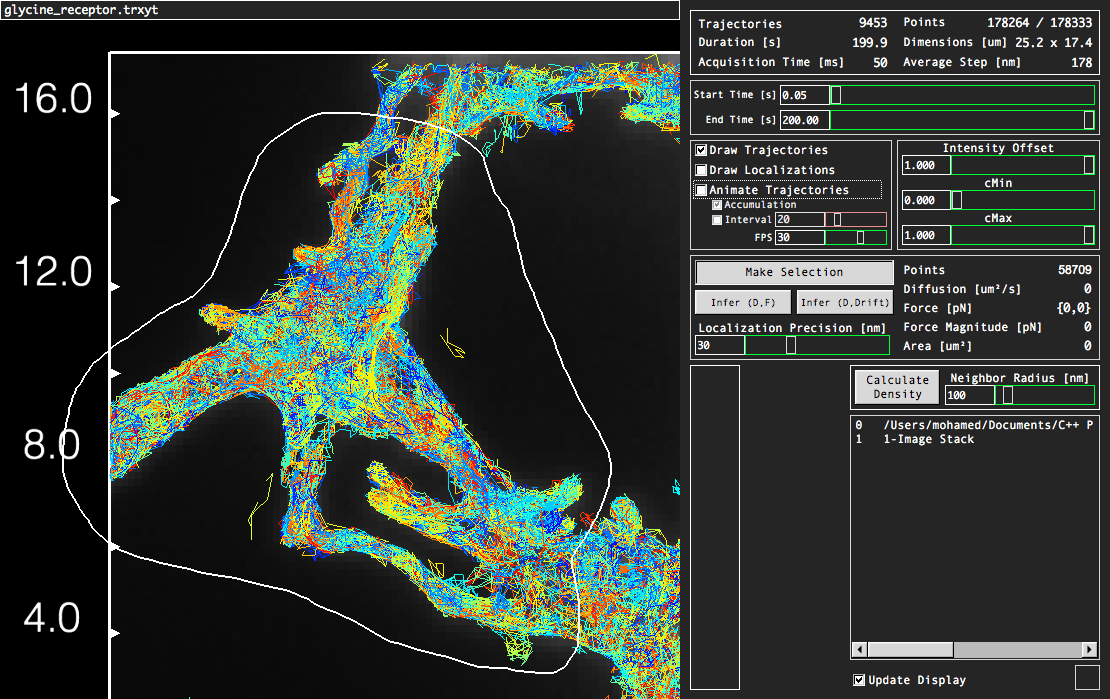}} \\
\end{tabular}

\begin{tabular}{ p{4.5cm} p{6cm} }
\multicolumn{2}{ p{\linewidth} }{\center\textbf{4.~Generate a Voronoi Mesh}} \\
\small In this example Voronoi meshing will be used on the user-selected region of the trajectories. Open the \textbf{Voronoi Meshing} interface from the \textbf{File > Inference > Meshing > Voronoi} menu. For this meshing mode, the number of zones must be predefined beforehand. \softwareName automatically selects an appropriate number based on the number of localizations in the loaded data set. Here we shall use the default value. Press the \textbf{Apply} button to generate the Voronoi mesh in only the custom selected region, which may take a couple minutes.
& \raisebox{-\height}{\includegraphics[scale=0.35]{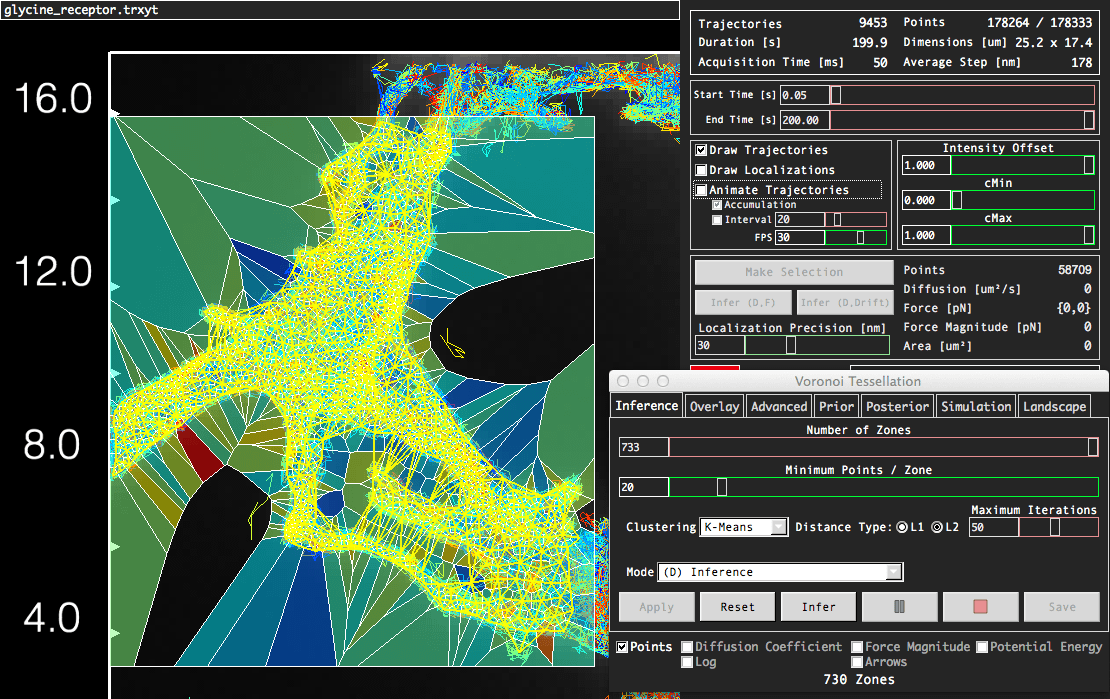}} \\
\end{tabular}

\begin{tabular}{ p{4.5cm} p{6cm} }
\multicolumn{2}{ p{\linewidth} }{\center\textbf{5.~Perform Inference}} \\
\small The yellow lines connecting each of the zones corresponds to which neighboring zones ``see each other'' (Described in Section \ref{sec:neighboring_zone_connections}). However, we shall be using \textbf{(D,F) Inference} in this example, which does not require zones to ``see each other''. Select the \textbf{(D,F) Inference} in the mode drop-down menu. Press \textbf{Infer} to perform the inference. Select \textbf{No} when prompted to compute the potentials.
& \raisebox{-\height}{\includegraphics[scale=0.35]{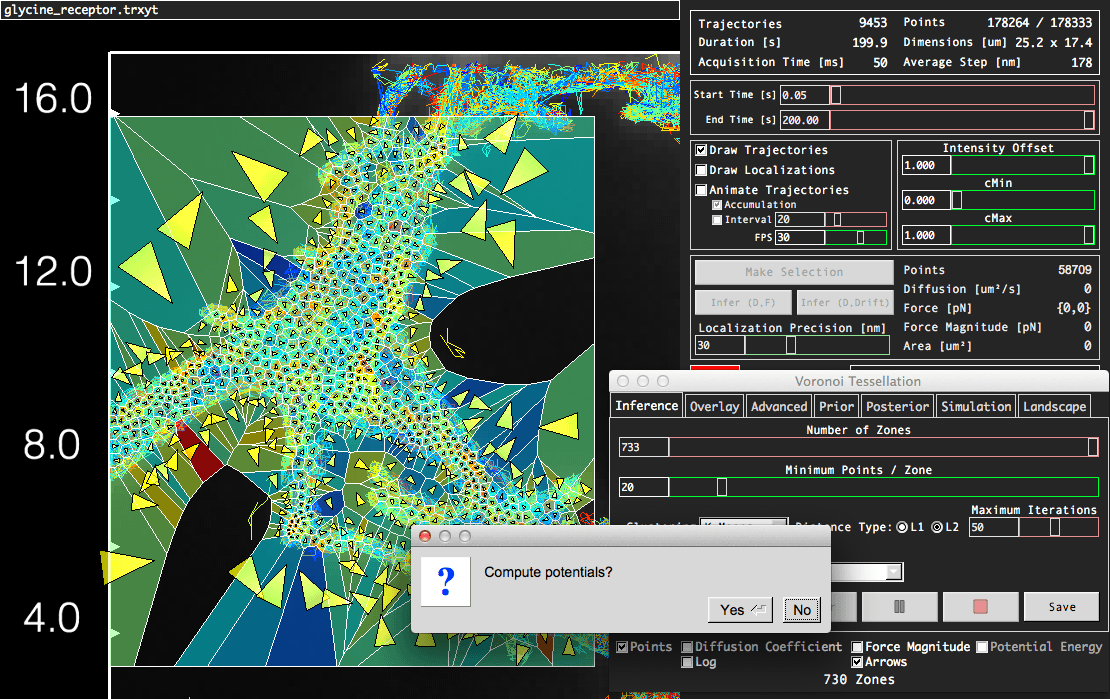}} \\
\end{tabular}

\begin{tabular}{ p{4.5cm} p{6cm} }
\multicolumn{2}{ p{\linewidth} }{\center\textbf{5.~Viewing Biases in Motion}} \\
\small Select the \textbf{Diffusion Coefficient} button in the bottom of the meshing interface. Create a box to zoom in on parts of the dendrites of the neuron. Directional arrows indicate regions where motion is systematically biased in a certain directions (animating the trajectories will confirm this).
& \raisebox{-\height}{\includegraphics[scale=0.35]{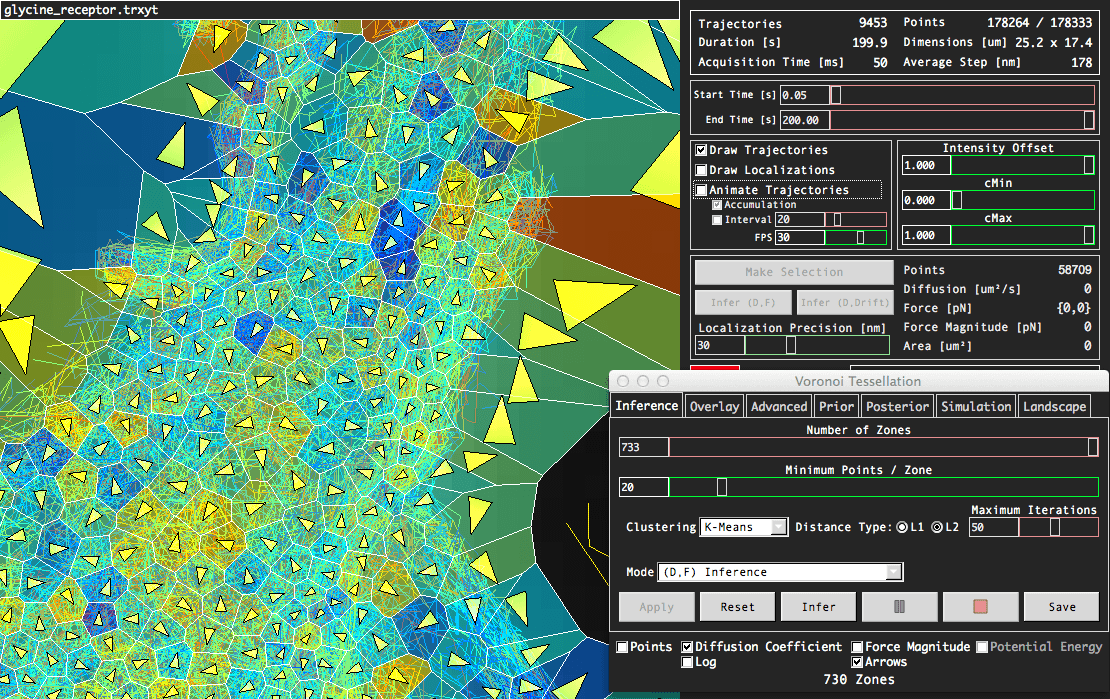}} \\
\end{tabular}

\clearpage



\clearpage

\section{Acknowledgments}

\subsection{Development Libraries}
Various freely-available development libraries were used in \softwareName. They are referenced below. In all cases, \softwareName conforms to all license agreements, and no modifications to the original authors' libraries are made.

\subsection{FLTK}
The \href{http://www.fltk.org/}{Fast Light Toolkit (\textbf{FLTK})} is the cross-platform graphical user interface library used in \softwareName. It was originally developed by Bill Spitzak.

\subsection{OpenGL}
\textbf{OpenGL} is the graphics library used for displaying graphics in \softwareName. It is maintained by the \href{http://www.khronos.org/opengl}{\textbf{Khronos Group}}.

\subsection{Libtiff}
\href{http://www.remotesensing.org/libtiff/}{\textbf{Libtiff}} is the library used for writing Tagged Image File Format files using \softwareName. It was originally developed by Sam Leffler at Silicon Graphics.

\subsection{TexFont}
\textbf{TexFont} is the textured font library used in \softwareName to display text in the display window. It was originally developed by Mark Kilgard.

\subsection{Clustering Algorithms}
Clustering algorithms (K-Means and H-Means) available for the voronoi tessellation modes are based on the C++ code by John Burkardt (original Fortran Code by David Sparks), available at: \url{http://people.sc.fsu.edu/~jburkardt/c_src/asa136/asa136.html}.

\clearpage
\section{Contact Information}
\label{sec:contact_information}

\begin{description}
	\item[Mohamed El Beheiry] (Developer) \href{mailto:mohamed.elbeheiry@gmail.com}{mohamed.elbeheiry@gmail.com}
	\item[Jean-Baptiste Masson] \href{mailto:jbmasson@pasteur.fr}{jbmasson@pasteur.fr}
\end{description}

\clearpage

\end{document}